\documentclass[aps,reprint,superscriptaddress]{revtex4-1}

\usepackage{graphicx}
\usepackage{physics}
\usepackage{braket}
\usepackage{hyperref}
\usepackage{amsfonts}
\usepackage{amssymb}
\usepackage{xcolor}
\usepackage{enumitem}
\graphicspath{{"../figures/"}}

\begin{document}

\title{
Optimal purification of a spin ensemble by quantum-algorithmic feedback
}

\author{Daniel M.\,Jackson\textsuperscript{1,*}}
\author{Urs Haeusler\textsuperscript{1,*}}
\author{Leon Zaporski\textsuperscript{1}}
\author{Jonathan H.\,Bodey\textsuperscript{1}}
\author{Noah Shofer\textsuperscript{1}}
\author{Edmund Clarke\textsuperscript{2}}
\author{Maxime Hugues\textsuperscript{3}}
\author{Mete Atat\"ure\textsuperscript{1,$\dagger$}}
\author{Claire Le Gall\textsuperscript{1,$\dagger$}}
\author{Dorian A.\,Gangloff\textsuperscript{1,$\dagger$}}

\noaffiliation

\affiliation{Cavendish Laboratory, University of Cambridge, JJ Thomson Avenue, Cambridge, CB3 0HE, UK}
\affiliation{EPSRC National Epitaxy Facility, University of Sheffield, Broad Lane, Sheffield, S3 7HQ, UK}
\affiliation{Universit\'e C\^ote d'Azur, CNRS, CRHEA, rue Bernard Gregory, 06560 Valbonne, France
\\ \ \\
\textsuperscript{*}\,These authors contributed equally to this work.
\\
\textsuperscript{$\dagger$}\,Correspondence should be addressed to: ma424@cam.ac.uk; cl538@cam.ac.uk; dag50@cam.ac.uk.
\\ \ \\
}

\begin{abstract}
Purifying a high-temperature ensemble of quantum particles towards a known state is a key requirement to exploit quantum many-body effects. An alternative to passive cooling, which brings a system to its ground state, is based on feedback to stabilise the system actively around a target state. This alternative, if realised, offers additional control capabilities for the design of quantum states. Here we present a quantum feedback algorithm capable of stabilising the collective state of an ensemble from an infinite-temperature state to the limit of single quanta. We implement this on $\sim$\,$50,000$ nuclei in a semiconductor quantum dot, and show that the nuclear-spin fluctuations are reduced $83$-fold down to 10 spin macrostates. While our algorithm can purify a single macrostate, system-specific inhomogeneities prevent reaching this limit. Our feedback algorithm further engineers classically correlated ensemble states via macrostate tuning, weighted bimodality, and latticed multistability, constituting a pre-cursor towards quantum-correlated macrostates.

\end{abstract}
\maketitle

\begin{figure*}[t!]
    \centering
    \includegraphics{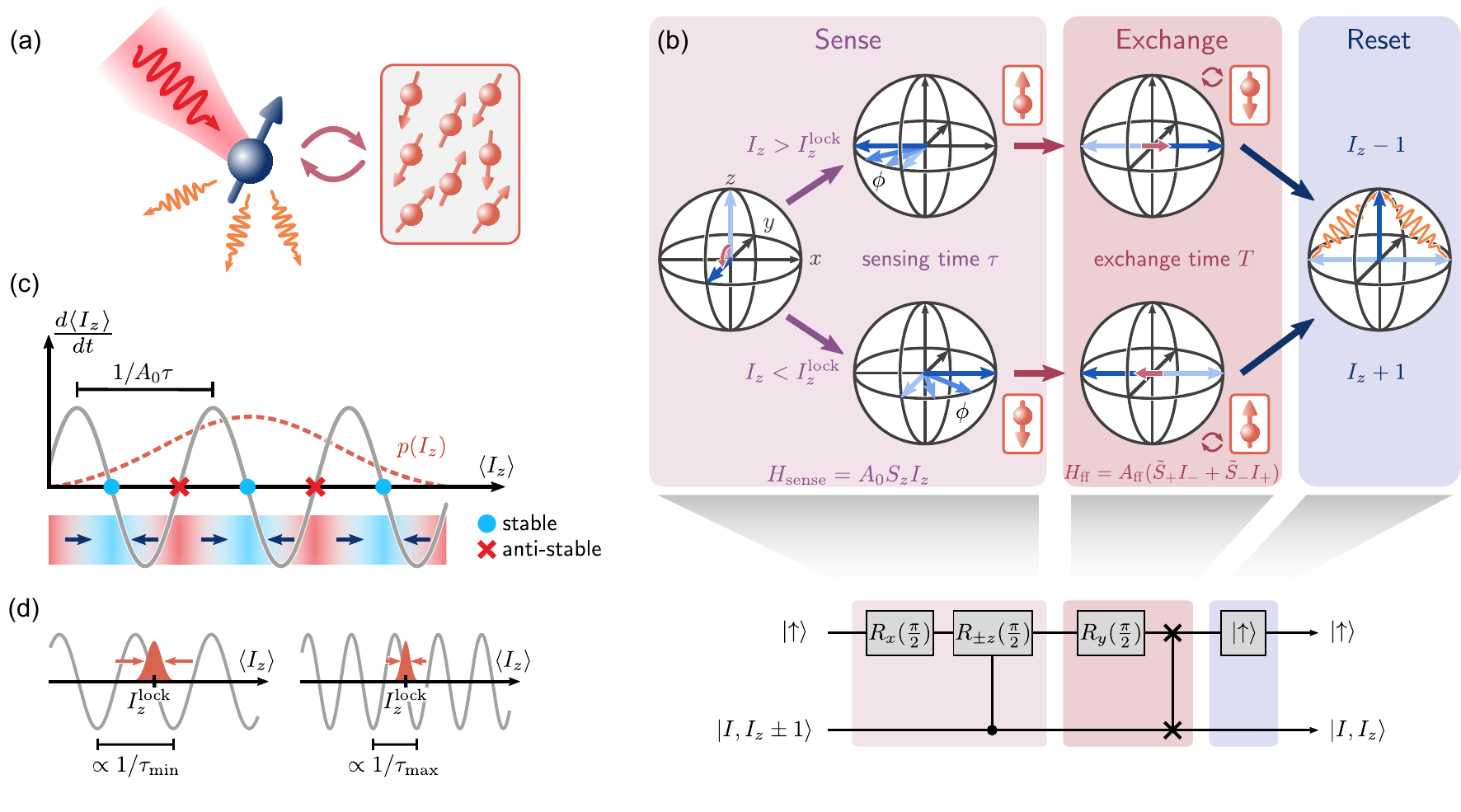}
    \caption{\textbf{Quantum feedback algorithm.} (a) Pulsed control (red shading) of a central spin (blue), homogeneously coupled to a spin ensemble (red), can purify the state of the system, removing entropy via spontaneous scattering (orange). (b) The feedback algorithm. (Top) Bloch sphere representation of the electron-spin evolution during one algorithm cycle. The upper (lower) path represents the case of a positive (negative) $\Delta I_z$. (Bottom) Quantum circuit representation of the feedback algorithm operating in the single-spin limit. The upper and lower rails represent respectively the central spin and the $\{\ket{I, I_z\pm1},\ket{I, I_z}\}$ states of the spin ensemble. (c) The rate of change $d\langle I_z\rangle /dt$ as a function of $\langle I_z\rangle$ (gray curve) on a coarse-grained evolution time $t\gg \tau + T$, displaying multiple (anti-)stable setpoints arising from the $1/A_0\tau$ periodicity of the sensing gate. The red dashed curve represents the initial thermal probability distribution. (d) To purify an initially broad distribution (left, red shaded area) to a narrow single mode centered around $I_z^\text{lock}$ (right, red shaded area), we increase the sensing time dynamically from $\tau_\text{min}$ to $\tau_\text{max}$.}
    \label{fig1}
\end{figure*}

\section{Introduction}
\label{sec_intro}

A controllable system of many interacting quantum objects hosts a phenomenally large Hilbert space which can serve as a versatile resource for both technological \cite{Monroe2021} and fundamental physics applications \cite{Abanin2019}. These range from realising multi-qubit registers for quantum information processing \cite{Taminiau2014,Debnath2016,Omran2019,Chekhovich2020} and storage \cite{Taylor2003b, Denning2019, Bradley2019}, to exploring collective phenomena such as superradiance \cite{Dicke1954,Wang2020} and discrete time crystal formation \cite{Choi2017,Zhang2017a}. Leveraging this resource requires reducing the ambient-condition entropy of such systems from that of a highly mixed thermal state to that of a pure state that reveals their quantum properties. Advances in cooling techniques have been transformative in achieving this goal in multiple physical platforms. Laser cooling of atomic gases and single trapped atoms \cite{Phillips1998} -- including Doppler, motional sideband-resolved, and spin-assisted Raman-based techniques \cite{Wineland1978,Monroe1995,Lee1996,Urvoy2019} -- as well as sideband cooling in electro-mechanical \cite{LaHaye04,Teufel2011}, opto-mechanical \cite{Groblacher2009}, and superconducting qubit systems \cite{Valenzuela2006} have been the trailblazers in this quest. In contrast to direct cooling, active stabilization at a target quantum state, in principle, also allows purification of a many-body state. Such an approach further enables the programmable preparation of non-equilibrium states and can be used to engineer designer ensemble distributions with varied many-body correlations. While the complex dynamics of highly degenerate many-body systems make stabilising a single \emph{microstate} very challenging, techniques to purify towards a single \emph{macrostate} remain highly desirable. Towards this end, the field of optimal quantum feedback control has developed an extensive toolbox \cite{Zhang2017} that can be exploited for quantum state engineering.

Feedback on a system comprises three elements \cite{franklin2002feedback}: sensor, controller and actuator. The sensor detects the current state of the system, a controller processes this information and tells the actuator how to correct the system towards a target state. In the case of quantum feedback, the control loop leverages quantum objects to enable sensing and actuation at the fundamental level of single quanta. A first example is measurement-based quantum feedback \cite{wiseman2010quantum}, which employs weak measurement of quantum observables to obtain classical information that is then processed by external electronics and used to control the actuator. This approach has been used for stabilising single qubit states \cite{Wang2001}, squeezed mechanical states \cite{Vinante2013}, photonic Fock states \cite{Zhou2012}, and mesoscopic spin squeezing \cite{Cox2016}, but limitations arise from measurement backaction and the rate-limiting classical electronics required. Going further, coherent quantum feedback \cite{Lloyd2000} overcomes these limitations by feeding quantum information directly from the sensor to the actuator without a measurement step. This enables autonomous stabilisation, and has been implemented in photonic \cite{Iida2012} and few-spin \cite{Nelson2000, Hirose2016} systems. Extending it to complex mesoscopic systems remains an open direction and, in this regime, a central spin coupled to a dense ensemble of nuclear spins serves as an ideal prototype. Techniques to stabilize the macrostate of this spin ensemble via the central spin, even far from the optimum of a single macrostate, have been a game changer: in gate-defined quantum dots (QDs), coherent sensing via an electronic proxy, combined with nuclear spin pumping, achieved a reduced-fluctuation, correlated state of two nuclear spin baths \cite{Bluhm2010}; whilst in optically active QDs autonomous feedback has opened a window into the many-body physics of the nuclei \cite{Latta2009,Ethier-Majcher2017,Gangloffeaaw2906}. Preparation of the large spin ensemble into a single macrostate, by achieving single-quanta level of control, remains an outstanding challenge.

In this work, we design an autonomous, time-sequenced quantum feedback algorithm capable of stabilising and engineering a mesoscopic spin system at the single macrostate limit. We apply this optimum feedback control to an optically active QD nuclear spin ensemble. This requires the deterministic correction of deviations from a target state at the level of a single quantum, which we achieve by exploiting the recent advances of sensing \cite{Jackson2021} and coherent control \cite{Gangloffeaaw2906} of single collective excitations in a nuclear ensemble -- nuclear magnons. Leveraging the coherence of these sensing and control processes, the central electron acts both as the sensor and actuator in a quantum feedback loop which, followed by a spin initialization step via optical pumping, removes entropy from the spin ensemble. We demonstrate a reduction by two orders of magnitude in the thermal fluctuations of a spin ensemble, and only a factor of 5 away from the fundamental quantum limit of single-spin fluctuations. We show that system-specific properties limit state stabilization of a single macrostate, but we verify through numerical simulation that this limit could be achieved in general. Finally, the control afforded by the use of quantum gates at each step allows to sculpt the feedback to generate highly non-trivial classically correlated states of the ensemble -- a precursor demonstration to generating quantum correlations.

\section{Results}

\subsection{3-step Feedback Algorithm}
\label{sec_algorithm}

The generality of our feedback algorithm allows it to be applied to a general central-spin \cite{Taylor2003} or central-boson \cite{Chen2015a} system (Fig.\,1a). We present each feedback step in general terms and realize their implementation with the physical system of a QD electron spin interfaced to $N \approx 50,000$ nuclear spins \cite{Urbaszek2013}. We parameterize this spin system by its collective state consisting of a total angular momentum $I$ and a polarization along the quantization axis $I_z \in [-I, I]$. In the case of a homogeneous one-to-all electron-nuclear coupling, the electron can change $I_z$ via single collective excitations, whilst $I$ is protected by symmetry \cite{Kozlov2007}. Thus the feedback actuated by the electron corrects one nuclear-spin deviation at a time, and in doing so, purifies the state of the nuclear-spin system. An external magnetic field of $3.5$\,T along the $z$-direction Zeeman-splits the electron spin, which we control with all-optical electron-spin resonance (ESR) allowing for fast multi-axis control \cite{Bodey2019} (supplementary materials section IA). The system Hamiltonian, expressed in a frame rotating at the ESR drive frequency $\omega$, is given by:
\begin{equation}
    H_0 = \delta S_z + \Omega S_x + \omega_\text{n} I_z+A_\text{c} S_z I_z+A_\text{nc}S_zI_x,
\end{equation}
where $S_i$ and $I_i$ are electron and nuclear spin operators respectively. The ESR detuning, Rabi frequency and nuclear Zeeman frequency are denoted $\delta=\omega_\text{e}-\omega$, $\Omega$, and $\omega_\text{n}$, respectively. The electron-nuclear coupling is enabled by the collinear ($A_\text{c}$) and the noncollinear ($A_\text{nc}$) constituents of the hyperfine constant. The collinear hyperfine term $A_\text{c} S_z I_z$ facilitates sensing by the electron since the ESR frequency is modified by a mean Overhauser field $A_\text{c}I_z$, and sensing of a single nuclear spin flip was recently shown in this system \cite{Jackson2021}. The electron is also the actuator thanks to the electron-nuclear exchange coupling enabled by the nuclear quadrupolar interaction, which reduces to a noncollinear hyperfine interaction $A_\text{nc}S_zI_x$ \cite{Huang2010, Hogele2012}. This coupling enables the injection of a single nuclear magnon \cite{Gangloffeaaw2906}, with a rate $f(I,I_z)A_\text{nc}$ that depends on the ensemble's total angular momentum $I$ and its polarization $I_z$ via the enhancement factor $f(I,I_z)\sim \mathcal{O}(\sqrt{N})$\cite{Dicke1954,Gangloff2020}.

 During feedback we employ sensing and actuation as sequential quantum gates. This allows us to leverage the available coherence for maximum fidelity gate operations at each step. With sensing and injection possible at the fundamental level of single quanta (nuclear spin flips) in the ensemble, the ideal feedback limit can be reached: detecting single-unit deviations from a target macrostate $I_z^\text{lock}$ and correcting them with exactly one unit. We first consider evolution during the feedback in the ideal, fully unitary case, except where dissipation is deliberately introduced. The feedback algorithm proceeds through the following three steps, which we visualize on the sequence of Bloch spheres and the corresponding quantum circuit in Fig.\,1b: 
\begin{enumerate}[leftmargin=*]
\item \textit{Sense}: A straightforward sensing mechanism for macrostate $I_z$ is a linear energy shift on the central spin: $H_\text{sense}=A_0S_zI_z$, with $A_0$ as a general coupling constant. The most efficient way to measure this energy shift is via Ramsey interferometry \cite{Degen2017}, with steps as follows. (i) The central spin starts spin-$\ket{\uparrow}$. (ii) An $R_x(\frac{\pi}{2})$-rotation places it in a coherent spin superposition in the Bloch equator. (iii) A free evolution time $\tau$ under $H_\text{sense}$ causes precession of the Bloch vector, whose projection along the $x$-direction will be the error signal for a deviation $\Delta I_z = I_z - I_z^\text{lock}$ from a target macrostate $I_z^\text{lock}$: $\expval{S_x}=-\frac{1}{2}\sin(2\pi A_0\Delta I_z\tau)$, which for a single spin flip $\Delta I_z = 1$ reaches a maximum at $\tau = 1/4A_0$. In the quantum circuit, this optimum evolution time corresponds to an $R_{\pm z}(\frac{\pi}{2})$-rotation conditional on the state of the spin system. (iv, optional) A final rotation $R_y(\frac{\pi}{2})$ may be required to convert this signal to an $\langle S_z\rangle$-polarisation, depending on the type of actuator gate that follows. Our example QD system, where $A_0 = A_\text{c}$, implements this sensing procedure naturally with $H_\text{sense}=\left(\delta +A_\text{c}I_z+A_\text{nc}I_x\right)S_z$. The ESR drive detuning $\delta \equiv -A_\text{c}I_z^\text{lock}$ determines the primary setpoint of the algorithm which feeds back on fluctuations $\Delta I_z = I_z-I_z^\text{lock}$. The transverse field $A_\text{nc}I_x$, which oscillates at $\omega_\text{n}$, contributes to sensing errors up to $\sqrt{N}A_\text{nc}/ A_\text{c}$, but can in principle be circumvented by an appropriate choice of B-field or $\tau$ (supplementary materials section VC2); it is an important consideration in determining the ultimate feedback limit.

\item \textit{Actuate}: Evolution under a flip-flop Hamiltonian $H_\text{ff}=A_\text{ff}(\Tilde{S}_+I_-+\Tilde{S}_-I_+)$ for a time $T$ converts the error signal $\langle \Tilde{S}_z \rangle=-\frac{1}{2}\sin(2\pi A_0I_z\tau)$ into a spin flip towards the target macrostate. Here $\Tilde{S}$ represents a simple basis rotation, reflecting the fact that some physical systems produce collinear flip-flop terms, where $\Tilde{S}_z=S_z$ (and the final $R_y(\frac{\pi}{2})$ is then required in the sensing step to make the error signal proportional to $S_z$), and others like our QD platform yield noncollinear terms, where $\Tilde{S}_z=S_x$. Evolution under $H_\text{ff}$ for a time $T=1/2f(I, I_z)A_\text{ff}$ performs a SWAP operation between the central spin and a single collective spin excitation. No measurement is made between steps $1$ and $2$, meaning the operation thus far is autonomous and reversible. In our QD system, we engineer the flip-flop Hamiltonian from the $A_\text{nc}S_zI_x$ term by driving the central spin at Hartmann-Hahn resonance \cite{Hartmann1962,Bodey2019}, $\Omega\approx\omega_\text{n}$, yielding $\Tilde{H}_\text{ff}=\Omega \Tilde{S}_z+\omega_\text{n}I_z-\frac{A_\text{nc}}{4}(\Tilde{S}_+I_-+\Tilde{S}_-I_+)$ \cite{Henstra2008}, where $A_\text{ff} = A_\text{nc}/4$.

\item \textit{Reset}: Up until this point the quantum feedback algorithm has corrected deviations from the setpoint, $\Delta I_z = I_z-I_z^\text{lock}$, by flipping a single spin within the ensemble entirely coherently, and therefore reversibly. To purify the spin ensemble further, we perform an irreversible reset operation on the central spin. In our QD system, this reset step is achieved by exciting the central spin to the charged exciton (trion) manifold with an optical pulse that incoherently pumps and re-polarizes the electron to state $\ket{\uparrow}$ with $>$\,$98\%$ probability. In doing so, we effectively transfer entropy from the spin ensemble to the photonic bath via the central spin, in analogy with heat-bath algorithmic cooling \cite{Park2016}.
\end{enumerate}

Applying the above algorithm repeatedly increases the purity of the spin bath and can, in principle, prepare a single $I_z$ macrostate. In practice, nuclear spin diffusion mechanisms, which are external to the feedback steps above, will compete with the feedback loop and limit the purity of the steady-state preparation. The approach to equilibrium under these competing effects can be gleaned qualitatively from a simple semiclassical rate equation governing the evolution of the mean value $\langle I_z \rangle$ \cite{Yang2013, Hogele2012} (supplementary materials section IV), valid over a coarse-grained evolution time $t \gg \tau + T$:

\begin{equation}
\frac{d\langle I_z \rangle }{dt} =\frac{-\sin(2\pi A_0\langle \Delta I_z \rangle \tau)}{\tau+1/2A_\text{ff}} -\Gamma_\text{d} \langle I_z \rangle.
\label{equ_dizdt}
\end{equation}
The first term is the rate at which the sensing, actuate, and reset gates together change $\langle I_z \rangle $ as a function of $\langle I_z \rangle$ -- it is the nonlinear dynamical function defining the feedback dynamics. The second term is a standard relaxation term capturing all spin diffusion mechanisms that relax a non-zero polarization $\langle I_z \rangle$ at a rate $\Gamma_\text{d}$. We see that in the low diffusion regime $\Gamma_d \ll A_\text{ff}/\left(1+ 2A_\text{ff}\tau\right)$, setting the sensing time to $\tau = 1/4A_0$ programs the feedback at its global optimal $T_0\times d\langle I_z \rangle/dt = -1$ for $\langle \Delta I_z \rangle = 1$, where $T_0=1/4A_0+1/2A_\text{ff}$ -- i.e. a fluctuation of one unit is fully corrected within a single algorithm cycle.

\begin{figure*}
    \centering\includegraphics{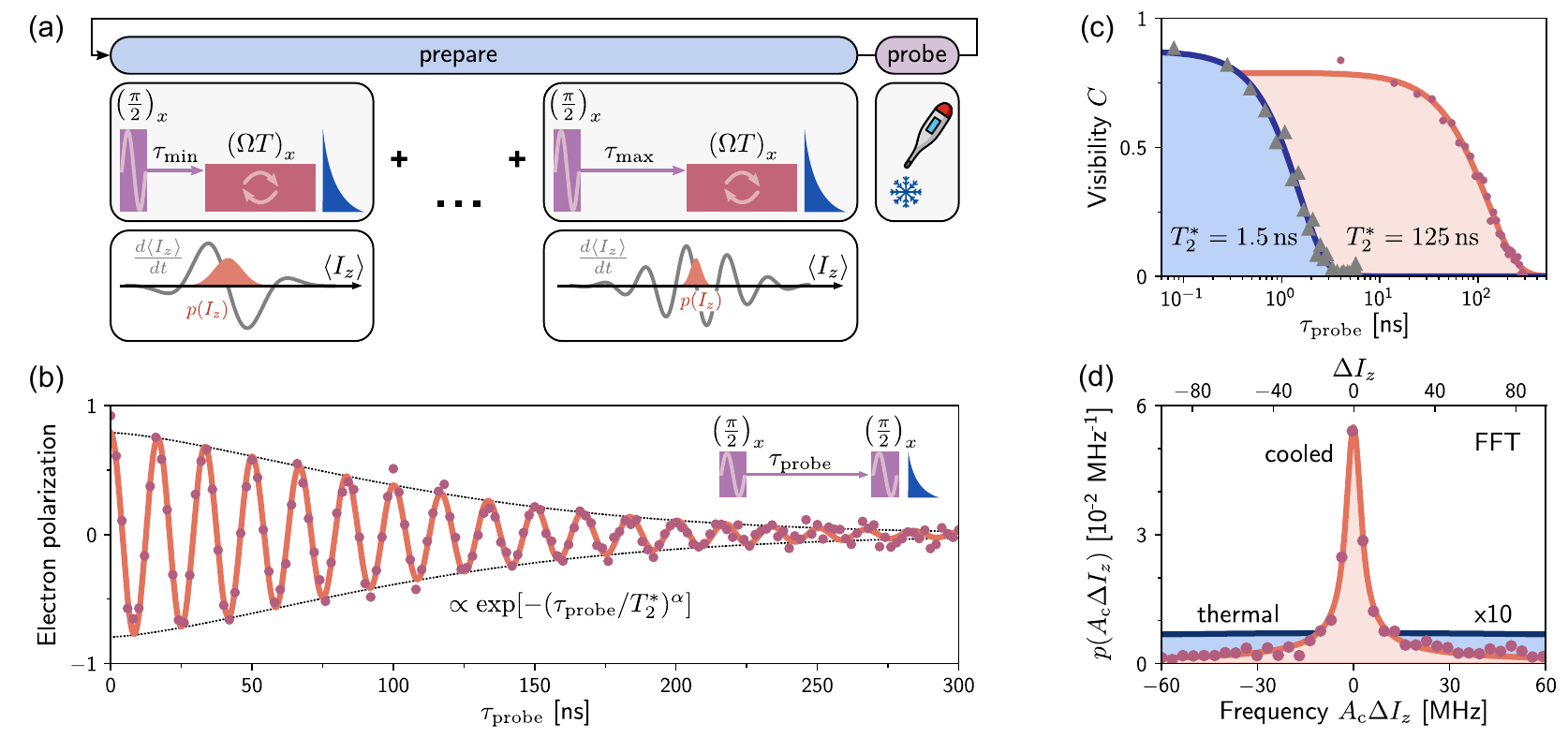}
    \caption{\textbf{Optimal quantum feedback.} (a) Feedback control sequence. We use $44$ elementary blocks of the algorithm. Each block has an $R_x(\frac{\pi}{2})$-rotation to initiate sensing (purple), an $R_x(\Omega T)$-pulse to actuate a spin flip via a Hartmann-Hahn resonance (red) and optical pumping for electron reset (blue). The sensing time is increased linearly over the $44$ blocks from $\tau_\text{min}$ to $\tau_\text{max}$, resulting in the schematic feedback curves (gray) beneath. (b) Electronic FID under optimal feedback. We alternate $\sim\,15\,\mathrm{\mu s}$ of feedback -- where $\tau_\text{min} = 30$\,ns, $\tau_\text{max} = 98$\,ns, $\Omega = 29$\,MHz, and $T = 86$\,ns -- with $\sim\,$2$\,\mathrm{\mu s}$ of probing $p(\Delta I_z)$ via Ramsey interferometry (inset), yielding a $60$-kHz repetition rate for single shots of the experiment. Each data point is an ensemble average measurement integrated for two seconds. The purple circles are the FID $\expval{S_z(\tau_\text{probe})}$ as a function of Ramsey delay $\tau_\text{probe}$. The FID oscillates at a frequency $\omega_\text{serr} = 60$\,MHz set by the phase, $2\pi\omega_\text{serr}\tau_\text{probe}$, which we add to the second Ramsey gate to make the process of fitting the envelope robust against small systematic detunings. The red curve is a phenomenological fit to a cosine with envelope $C(\tau_\text{probe}) = \exp\left[-\left(\tau_\text{probe}/T_2^*\right)^\alpha\right]$ (dotted curve), where $T_2^*=125(4)$\,ns and $\alpha=1.46(9)$ (c) Triangles (circles) are the FID envelopes resulting from a nuclear spin ensemble without (with) the application of our optimized feedback algorithm. Fitting these data with the blue and red $C(\tau_\text{probe})$ curves yields $T_2^*=1.52(5)$\,ns, $\alpha=1.60(12)$ and $T_2^*=125(4)$\,ns, $\alpha=1.46(9)$ respectively. (d) Fourier transform of data (circles) and FID envelopes (curves) from Fig.\,2c, yielding explicitly the probability distribution $p(A_\text{c}\Delta I_z)$ for the purified (thermal) ensemble shown in red (blue).}
    \label{fig2b}
\end{figure*}

\begin{figure*}
    \centering\includegraphics{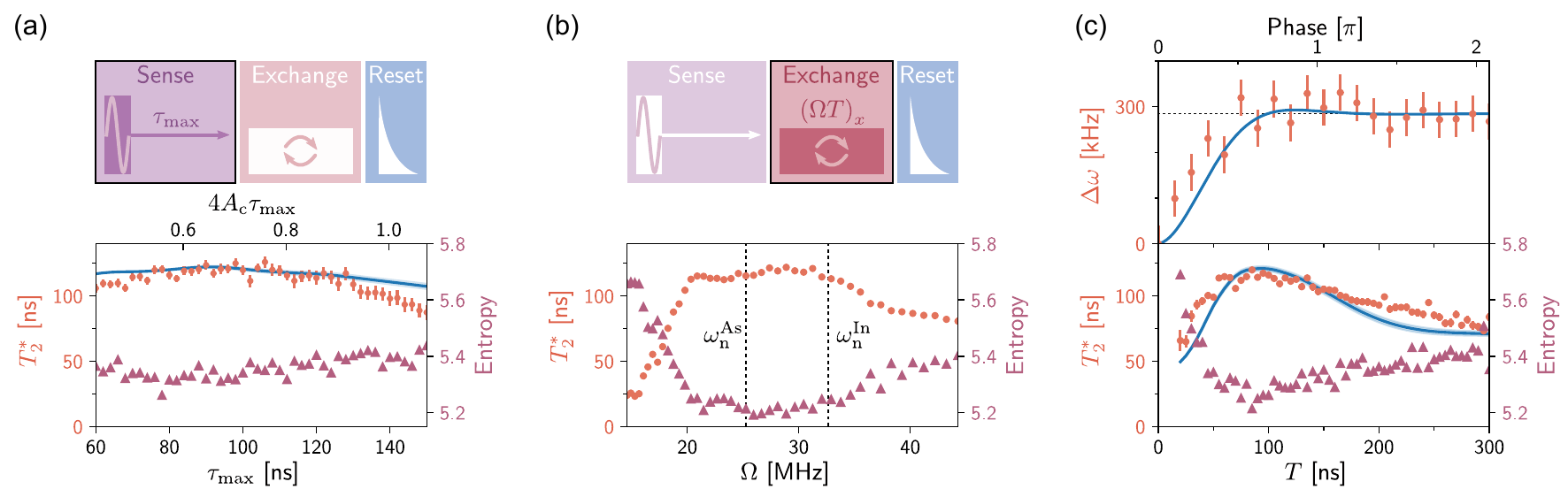}
    \caption{\textbf{Optimising feedback parameters.} (a) (Top) Varying sensing gate $\tau_\text{max}$ in the feedback sequence. (Bottom) $T_2^*$ (orange circles, left vertical axis) and entropy $S_p$ (purple triangles, right vertical axis) versus $\tau_\text{max}$ for $\Omega = 29$\,MHz and $T = 86$\,ns. The blue curve is a fit of $T_2^*$ versus $\tau_\text{max}$ obtained from numerical simulation, with the blue shading as a $68\%$ confidence interval. (b) (Top) Varying actuate gate $\Omega$ and $T$ in the feedback sequence. (Bottom) $T_2^*$ (orange circles, left vertical axis) and entropy $S_p$ (purple triangles, right vertical axis) versus $\Omega$ for $\tau_\text{max} = 100$\,ns and $T = 86$\,ns. Vertical dashed lines indicate the position of the Hartmann-Hahn resonances for arsenic and indium. (c) (Top) ESR shift from ultra-precise Ramsey measurement \cite{Jackson2021} as a function of actuate gate duration $T$ (orange circles), fitted (blue curve) with the same microscopic model as for the feedback (supplementary materials section V), shows a maximum population transfer at $T \sim114$\,ns. The fit parameters of this model are $A_\text{c}=0.63(2)$\,MHz, $A_\text{nc}=140(13)$\,kHz and a pure nuclear dephasing rate $\Gamma=6(2)$\,MHz. (Bottom) $T_2^*$ (orange circles, left vertical axis) and entropy $S_p$ (purple triangles, right vertical axis) versus $T$ for $\Omega = 31$\,MHz and $\tau_\text{max} = 100$\,ns. The blue curve is a fit of $T_2^*$ versus $T$ obtained from numerical simulation, with the blue shading as a $68\%$ confidence interval. The parameters for the simulated $T_2^*$ curves in (a) and (c) are $A_\text{c}=0.63$\,MHz, $A_\text{nc}=156(2)$\,kHz and $\Gamma=6$\,MHz.}
    \label{fig2c}
\end{figure*}

Figure 1c shows the curve described by Eq.\,\ref{equ_dizdt}, on which we have highlighted the stable points of the feedback dynamics defined by a zero crossing and a negative slope of $d \langle I_z\rangle/dt$. A key feature is the existence of multiple stable points split by $1/A_0\tau$ in the $\langle I_z \rangle$ phase space. This arises intuitively from the $2\pi$-periodic temporal phase acquisition during sensing, meaning the feedback does not distinguish between points in the phase space where $A_0\Delta I_z\tau\in \mathbb{Z}$. The splitting between stable points is effectively the capture range for each stable point in the $I_z$ phase space -- all fluctuations within this capture range are shepherded back towards the same stable point. Polarization fluctuations are described fully by the probability distribution of macrostates $I_z$: $p(I_z)=\langle I_z| \text{Tr}_\text{e}(\rho)|I_z\rangle$, where we trace the full system's density matrix $\rho$ over the central spin ($e$). In this picture, a physical system with an initial state at large temperature (that is, all microstates are equiprobable) exhibits a broad initial $p(I_z)$ distribution (Fig.\,1c), with variance $\langle \Delta I_z^2 \rangle \sim N$. Adapting the feedback capture range to the system's initial state is thus critical to lock the system to a desired stable point $I_z^\text{lock}$; a capture range narrower than the typical width of $p(I_z)$ splits the ensemble into multiple stabilized modes. However, extending the capture range by shortening the sensing duration results in a reduced feedback strength, as per Eq.\,\ref{equ_dizdt}. To resolve this tension, we vary the sensing time dynamically, changing it from sequence to sequence, such that the first in the series $\tau_\text{min} \sim1/4A_0\sqrt{N}$ has a capture range sufficient for a thermal state and the last in the series $\tau_\text{max} \sim1/4A_0$ optimally corrects single-spin fluctuations (Fig.\,1d).

\subsection{An ultra-narrow QD nuclear ensemble}
\label{sec_QD}

Figure 2a shows the full control sequence we employ in our QD system for feedback, where we have $15\,\mu\text{s}$ of nuclear-state purification, consisting of $44$ elementary units of the algorithm. Between each unit, we increase the sensing time linearly from $\tau_\text{min}=30$\,ns to $\tau_\text{max} \leq 150$\,ns, and Fig.\,2a shows the effective feedback curves at each step. In contrast to Fig.\,1 the $d\langle I_z\rangle/dt$-curves have an envelope function corresponding to the finite bandwidth of our feedback implementation. This width is dictated by the electron-nuclear coupling rate during actuation $\frac{1}{4}A_\text{nc}\sqrt{N/2}\sim 2\,\mathrm{MHz}$, which restricts efficient polarization transfer to $\Delta I_z$-fluctuations satisfying $\sqrt{\omega_\text{n}^2+(A_\text{c}\Delta I_z)^2}-\omega_\text{n}\lesssim 2\,\mathrm{MHz}$, thereby defining a full bandwidth $|A_\text{c}\Delta I_z| \approx 20$\,MHz. The preparation step is followed by a probe step to determine the characteristic width of the probability distribution $p(\Delta I_z)$, which measures how close the prepared system is to an ideal single-macrostate $I_z^\text{lock}$. To measure $p(\Delta I_z)$, we again use the electron's sensing capability; specifically we use Ramsey interferometry (Fig.\,2b). An $R_x(\frac{\pi}{2})$ gate applied to an initialized electron spin followed by a wait time $\tau_\text{probe}$ leads to phase accumulation which senses the mean field $A_\text{c}I_z$. This phase is mapped to electron population $\rho_{\uparrow\uparrow}$ with a final $R_x(\frac{\pi}{2})$ gate. We use a second Ramsey measurement but with a final $R_{-x}(\frac{\pi}{2})$ gate to obtain a calibrated measurement of electron polarisation (supplementary materials section IB). By repeating $\mathcal{O}(10^5)$ such pump-probe measurements over a few seconds, we obtain the ensemble average electronic evolution over $p(A_\text{c}\Delta I_z)$, namely the free induction decay (FID) of the electronic spin. Figure 2b shows the electron polarization as a function of the probe time $\tau_\text{probe}$, that is, the FID following preparation with the feedback sequence of Fig.\,2a, with visible coherence extending to $300$\,ns. 

Fitting the FID with a stretched exponential envelope $C(\tau_\text{probe}) = \exp\left[-\left(\tau_\text{probe}/T_2^*\right)^\alpha\right]$, where $\alpha>0 $ is a free parameter, we find an electronic coherence time $T_2^*=125(4)$\,ns (Fig.\,2c, red curve) -- the longest reported to date in this system for any electronic qubit \cite{Ethier-Majcher2017,Huthmacher2018}. This is an improvement by a factor of $83$ relative to the FID taken with a probe measurement of the system without a preparation step. Indeed, a probe measurement of the thermal ensemble at the ambient temperature for our experiments ($4$\,K) yields a coherence time $T_2^*=1.52(5)$\,ns (Fig.\,2c, blue curve). The envelope function $C(\tau_\text{probe})$ contains all the information about the nuclear macrostate distribution $p(A_\text{c}\Delta I_z)$ via a Fourier transform \cite{Onur2018}. Figure 2d shows $p(A_\text{c}\Delta I_z)$ for the thermal ensemble (red curve), whose full-width at half maximum (FWHM) is approximately $330$\,MHz, in agreement with $A_c\sqrt{5N/4}$ for $A_c = 0.63(2)$\,MHz and $N=4.9(4)\cdot 10^4$ (supplementary materials section IIB), and representing a distribution over $\sqrt{N} = 220$ macrostates. By comparison, the cooled distribution has a FWHM of $6$\,MHz, equivalent to a probability distribution over approximately $10$ macrostates. We note here that whilst these macrostates are not necessarily resolved owing to a non-uniform one-to-all hyperfine coupling, the ratio of the width of the distribution to the average $A_c$ is still representative of an effective number of macrostates contained within the distribution.


\subsection{Optimising feedback in a QD system}

We arrive at this global optimal of $10$ macrostates by tuning the constitutive variables of the feedback algorithm, namely the maximal sensing time $\tau_\text{max}$, the ESR Rabi frequency $\Omega$, and the duration of the actuate gate $T$. The feedback performance is characterized using two different metrics: the electron dephasing time $T_2^*$, which we wish to maximise, and an information entropy $S_p$, which we wish to minimise. The characteristic dephasing time, $T_2^*$, is obtained as in Fig.\,2 by fitting an envelope $C(\tau_\text{probe}) = \exp\left[-\left(\tau_\text{probe}/T_2^*\right)^\alpha\right]$ to the FID data. This $T_2^*$ works well to capture the effect of purifying the spin ensemble as long as the nuclear-spin distribution remains in a single mode, but fails otherwise. The information entropy $S_p$ of $p(A_\text{c}\Delta I_z)$ is the limiting density of discrete points (supplementary materials section IIA), where $p(A_\text{c}\Delta I_z)$ is the Fourier transform of the FID data. This entropy measure, which extends the notion of Shannon entropy to probability density functions, is a complete, model-independent measure of our data which does not require fitting. It captures the purification in $I_z$, which we treat as a classical noise source, irrespective of the underlying $I$-degeneracy.

Figure 3a shows the electron dephasing time $T_2^*$ (orange circles) and the information entropy $S_p$ (purple circles) as a function of the maximum sensing time $\tau_\text{max}$, in a linear sweep from $\tau_\text{min}=30$\,ns to $\tau_\text{max}$ (as in Fig.\,2a). In section IIIA of the supplementary materials we verify that this $\tau_\text{min}$ is sufficiently short to ensure a capture range large enough to stabilize the ensemble distribution around a single mode. Our actuate gate has a finite bandwidth ($\sim 20$\,MHz) corresponding to a sensing time of approximately $30$\,ns, and it is not necessary to use shorter sensing times because nuclear diffusion brings any initial nuclear state to this relatively broad $20$\,MHz window. We find the optimum $\tau_\text{max}$ at approximately $90$\,ns, where the maximum value of $T_2^*$ and the minimum value of $S_p$ coincide. This value is significantly shorter than the theoretical optimum of $1/4A_\text{c}$\,$\approx$\,$400$\,ns given a fitted hyperfine constant $A_\text{c}$\,$\approx$\,$0.63(2)$\,MHz (supplementary materials section VF). To understand this deviation, we simulate numerically the effect of our cooling algorithm on $N=49,000$ spin-$1/2$ nuclei after mapping the problem to a smaller Hilbert space for computational feasibility. For tractability we further employ exact diagonalization for the unitary evolution during sensing and actuate gates \cite{Kozlov2007}, whilst Kraus operators capture relaxation and dephasing via amplitude- and phase-damping channels (supplementary materials sections VB and VC). For independently measured values of $A_\text{c}=0.63$\,MHz and pure nuclear dephasing rate $\Gamma=6$\,MHz we obtain good agreement between simulation (blue curve) and measurements, for $A_\text{nc}=156(2)$\,kHz and, in line with previous work \cite{Gangloffeaaw2906, Jackson2021}, a subset of nuclei ($\sim$\,$21,000$) partaking in the actuate gate (supplementary materials sections VF). We find that electron dephasing during sensing caused by fluctuations of the finite transverse field $A_\text{nc}I_x$ (see eq.\,1), which we capture in our simulation as a semi-classical magnetic field noise of amplitude $\sim$\,$\sqrt{N}A_\text{nc}$, explains quantitatively the observed optimal sensing time. We note that this limitation is specific to the non-collinear hyperfine coupling in our QD platform, not the general feedback algorithm, and we verify that when this transverse noise is removed the optimum indeed occurs at the expected time $1/4A_\text{c}$ (supplementary materials section VG).

Figure 3b shows the electron dephasing time $T_2^*$ (orange circles) and the information entropy $S_p$ (purple circles) as a function of the ESR Rabi frequency $\Omega$ used to activate the flip-flop exchange gate. With a fixed drive time of $T=86$\,ns we see an optimum Rabi frequency of $\Omega\approx29$\,MHz. This is in close agreement with our theoretical expectation that the strongest feedback occurs when the actuate gate consists of an ESR drive on Hartmann-Hahn resonance. This QD system exhibits two such resonances at the corresponding Zeeman energies of two nuclear species: arsenic at $\omega_\text{n}^\text{As}=25$\,MHz and indium at $\omega_\text{n}^\text{In}=33$\,MHz. Thanks to the quadrupolar-induced, few-megahertz inhomogeneous broadening of the nuclear Zeeman levels \cite{Stockill2016}, the optimal ESR Rabi frequency occurs around the average of the two resonances.

In seeking to optimize the duration of the activate gate, we expect the optimum setting to leverage the available coherence during the electron-nuclear interaction to achieve a maximal fidelity swap gate. As a first step, we thus characterize the coherence of the electron-nuclear interaction in a direct measurement of the electron-nuclear polarization transfer during the exchange gate, as shown in the top panel of Fig.\,3c. Following a preparation sequence as in Fig.\,2a, the probe step alone is replaced with an actuate gate of duration $T$, which swaps polarization from the electron to a single nuclear magnon, followed directly -- without an intervening projective measurement -- by an ultra-precise Ramsey measurement \cite{Jackson2021} that measures the $A_\text{c}$-scale ESR shift induced by the magnon (orange circles). Fitting this data with the same modeling approach used for the actuate gate of the algorithm (blue curve) confirms that the electron-nuclear exchange is coherent, and has a maximum population transfer at $T\approx 114$\,ns. Furthermore, this data yields a direct measure of the collinear hyperfine constant \cite{Jackson2021}, $A_\text{c}=0.63(2)$\,MHz and pure nuclear dephasing rate $\Gamma=6(2)$\,MHz, which we use to constrain the model for the feedback algorithm (supplementary materials sections VF).

Figure 3c (bottom panel) shows the electron dephasing time $T_2^*$ (red circles) and the information entropy $S_p$ (purple circles) as a function of the duration $T$ of the actuate gate, where we find the optimum at $T\approx86$\,ns. The model values $A_\text{nc}$ and the number of nuclei partaking in the actuate gate for both the $T_2^*$-dependence and electron-nuclear polarization transfer agree very closely (supplementary materials sections VF). From these values, we would predict an optimum for feedback at the $\pi$-time of the interaction $T$\,$\sim$\,$2/(A_\text{nc}\sqrt{N/2})$\,$\sim$\,$130$\,ns. The approximate agreement between our measured optimum and this simple theoretical estimate confirms that the feedback is optimal close to the maximum achievable fidelity of the swap operation. Furthermore, our model informs us that our measured optimum is modified from the theoretical optimum by an optically induced electronic spin relaxation process whose rate is proportional to the power of the incident laser light enabling spin control \cite{Bodey2019}. Under our experimental conditions, this electronic relaxation rate is such that the electron spin is close to completely depolarized when the electron-nuclear exchange reaches its maximum at the $\pi$-time of the interaction. This can also be seen in the sensing measurement of Fig.\,3c, top panel. When we turn on the actuate gate on an unpolarized electron spin, the effect of the gate is to diffuse the nuclear state away from the lockpoint. Thus the cooling performance can improve by reducing the drive time $T$, which in the vicinity of the $\pi$-time reduces the driven diffusion significantly more than the electron-nuclear polarization transfer. Finally, we verify separately that removing altogether this relaxation process from our numerical simulations indeed returns the optimal drive time to the $\pi$-time $\sim 2/(A_\text{nc}\sqrt{N/2})$ (supplementary materials section VG).

\begin{figure*}
    \centering\includegraphics{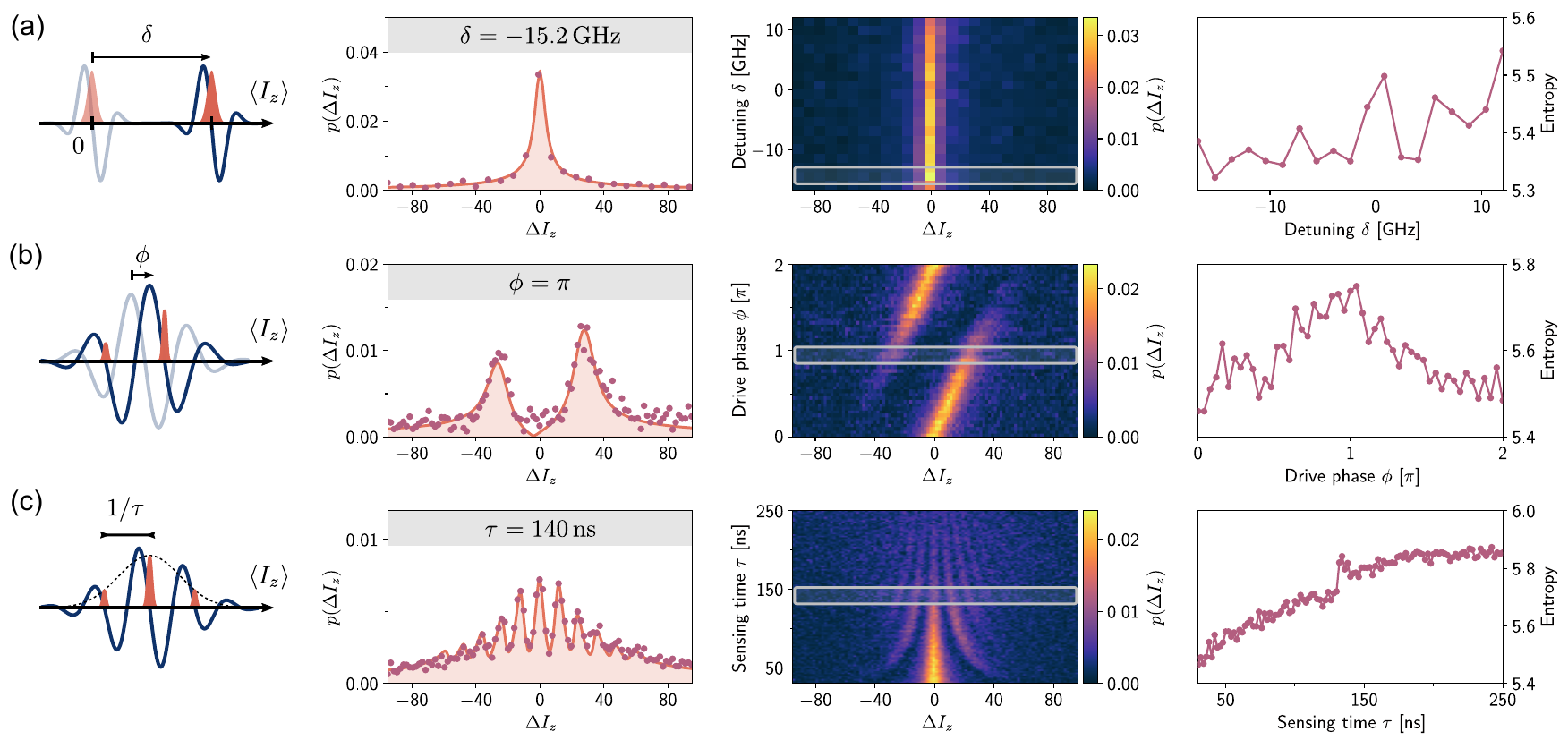}
    \caption{\textbf{Engineering spin-ensemble distributions.} From left to right: sketch of how the feedback curves (solid blue) are programmed to engineer $p(I_z)$ (orange shaded areas); plots of the measured probability density functions $p(A_\text{c}\Delta I_z)$ (purple circles) and their fits (solid curves) at a salient point in the relevant parameter space; full experimental two-dimensional map, from which these salient slices (gray boxes) are taken, showing continuous tuning over the parameter space; information entropy, $S_p$, (purple circles) of the $p(A_\text{c}\Delta I_z)$ distribution versus the relevant tuning parameter. (a) Varying $\delta$ sets the central lockpoint $I_z^\text{lock}$ of the feedback. The extrema of the mean field, $A_\text{c}I_z^\text{lock}$, possible with our feedback algorithm are $+12.0$\,GHz, and $-15.2$\,GHz, where the FWHM of $p(A_\text{c}\Delta I_z)$ are measured as $9.7$\,MHz and $5.8$\,MHz, respectively. (b) Tuning the drive phase $\phi$ allows continuous variation between mono- and bi-modal feedback as seen in the $\delta$-$\Delta I_z$ map. (c) Using a single sensing time $\tau$ during feedback determines a well-defined steady-state lockpoint splitting. Varying $\tau$ in the two-dimensional map shows we can tune the number of modes in the probability distribution.}
    \label{fig3} 
\end{figure*}

\subsection{Engineering classical correlations}
\label{sec_corr}

Our feedback algorithm allows further engineering of the spin distribution by tuning the feedback curve to host a set of desired locking points -- single or multiple. Within the dynamical landscape experienced by the collective spin state $I_z$, this creates programmable trapping points for the spin ensemble. This is made possible by direct control over parameters that define the sense and actuate quantum gates, namely the ESR detuning $\delta$, the relative phase $\phi$ between sense and actuate gates, and the sense time $\tau$.

Using the ESR detuning $\delta$ as a tuning parameter, Fig.\,4a demonstrates precise control over the mean of the $I_z$ distribution by translating the lockpoint $I_z^\text{lock}$ from that of an unpolarized ensemble ($I_z^\text{lock} = 0$). Our polarising sequence steps the ESR detuning as a function of time $t$, $\delta(t) = -A_\text{c}I_z^\text{lock}$, in steps of $20$\,MHz thereby stepping the setpoint by $\sim 30$ spin flips, which is detected by the sensing step as a non-zero error signal. At each new value of $\delta(t)$ we perform $\sim 2500$ complete cooling sequences to fully correct this error signal. Doing so slowly enough relative to the typical feedback rates and in steps much less than the width of the feedback capture range allows nuclear polarization to build up, leading to an average \emph{dragging} process equivalent to dynamic nuclear polarization \cite{Hogele2012}. Crucially, we find that the feedback remains stable over a range of fractional nuclear polarization of approximately $-20\%$ to $+30\%$, as estimated from the measured Overhauser shift by assuming equal fractional polarization for each species and an indium concentration of $0.5$ \cite{Stockill2016}. This is evidenced by a nuclear spin distribution whose width remains within a factor of $2$ of the optimal and whose information entropy remains largely unchanged over this domain (Fig.\,4a, right panel). The dynamic locking range, defined as the range of achieved lockpoints over the width of the distribution, corresponds to well over 1,000 distinct accessible macrostates.

Using the relative phase $\phi$ between the sense and the actuate gates as a tuning parameter, we can transform the probability distribution $p(A_\text{c}\Delta I_z)$ from a single-mode to a bi-modal distribution, with a programmable frequency-mode splitting equal to the inverse sensing time $1/\tau$. This is done by modifying the first pulse of the algorithm $R_x(\frac{\pi}{2})$ to have instead a phase $\phi$ relative to the $R_x(\Omega T)$ actuate gate. This alters the phase of the error signal acquired during sensing, such that $\expval{S_x}=-\frac{1}{2}\sin(2\pi A_0\Delta I_z\tau-\phi)$. Its effect on the feedback curve is shown in Fig.\,4b (left panel), where given the envelope function associated with our feedback bandwidth, the phase $\phi$ is equivalent to a carrier-envelope phase. By changing this phase, we can tune the relative weight of the two possible modes of the distribution, as shown in the two-dimensional map of Fig.\,4b, for values of $\phi$ from $0$ to $2\pi$. As can be seen from the right panel of Fig.\,4b, the entropy is maximized for $\phi = \pi$, where there exists an equal weighting between the two modes. This particular point can be seen as a balanced classical superposition of two $I_z$ states separated by $\sim 55$ nuclear spin flips, each with a width of a few spins. Detecting this classical signal is a first step towards the investigation of a Schr\"odinger kitten state \cite{Ourjoumtsev2006}: in this case a quantum superposition of states of mesoscopic nuclear polarisation.

Finally, using the sense time $\tau$ as a tuning parameter, we create multi-modal frequency-comb-like spin distributions. To do so, we apply a fixed sense time $\tau$ during our feedback sequence (Fig.\,2a), which tunes a correspondingly fixed capture range, set by the spacing $1/A_0\tau$ between two lockpoints within the feedback function. When this capture range is smaller than the width of the initial nuclear distribution, the nuclear state $I_z$ probabilistically falls into any one of the lockpoints contained within the initial distribution, as shown in Fig.\,4c (left panel). In Fig.\,4c, we demonstrate this for $\tau = 40$\,ns to $\tau = 250$\,ns, achieving from $3$ to $11$ simultaneous modes of the spin distribution, respectively. Strikingly, at the long sensing times $\tau \gtrsim 140$\,ns, the initial state is frozen into $\gtrsim 9$ modes, each with a width of $4$\,MHz or approximately $6A_\text{c}$, which is well below the $6$\,MHz width measured for the optimal single-mode preparation. The increased feedback strength at these long sensing times gets the system closer to its ultimate narrowing limit at the expense of populating other nearby modes. Whilst the individual modes of the distribution become narrower, the information entropy (Fig.\,4c right panel) of the overall distribution increases with sensing time as the distribution is spread over an increasingly large phase space. These multi-mode spin distributions can be seen as a freezing of the initial state onto a one-dimensional lattice of points in the phase space of nuclear polarisation, which we call latticed multistability. Due to a finite probability of hopping between lattice sites, integrating over a sufficiently long time yields the observed ensemble average. With added phase coherence between sites, this lattice could simulate a one-dimensional quantum walk where the lattice spacing and depth can be tuned with the sensing time and the amplitude of electronic coherence during sensing, respectively.

Such control over the modes of a spin distribution via a tailored dynamical landscape is equivalent to generating classical correlations of a collective spin state on demand. It serves as a precursor to phase-coherent state preparation within this manifold of stable points that would, in principle, result in programmable quantum superposition states of this collective spin.

\section{Discussion}

We presented a new quantum-algorithm approach to feedback on general central spin or boson systems, which can stabilize fluctuations of the ensemble down to a single quantum. Applying this feedback to a QD nuclear-spin ensemble enabled an $83$-fold reduction in spin fluctuations, and the engineering of non-trivial classically correlated nuclear states. Whilst we show the generation of a classical superposition of two $I_z$ states separated by tens of spin flips, the dissipative reset of the electron within each feedback cycle precludes any quantum coherence between the two. Thus a natural extension of this work will be the design of a modified control sequence, where the electron is disentangled before the reset step, maintaining nuclear coherences. A technique of this kind and with the level of control we have demonstrated herein could generate quantum superpositions of two or more mesoscopic nuclear states. A future direction also includes investigating the limit cycle during feedback operation at the single-spin level: applying the algorithm to an initial pure state $\ket{I_z}$, the nuclear state would cycle between $\ket{I_z}\bra{I_z}\leftrightarrow \frac{1}{2}(\ket{I_z-1}\bra{I_z-1}+\ket{I_z+1}\bra{I_z+1})$ sub-harmonically with a periodicity twice that of the algorithm. The appearance of this limit cycle could be a witness for the dissipative preparation of a well-defined total angular momentum state \cite{Gangloff2020}, $I$, constituting the initialization of a pure Dicke state in this central-spin system. 

\section{Materials and Methods}

\subsection{Sample}
The heterostructure of the wafer used in this work, which has been used in previous studies \cite{Stockill2016, Ethier-Majcher2017, Huthmacher2018, Bodey2019, Gangloff2020}, consists of an InGaAs QD layer integrated into a Schottky diode structure which allows to control the charge state of the QD. At the bottom of the heterostructure is a distributed Bragg reflector to improve photon emission from the top surface. The photon collection is further enhanced to 10\% at the first lens with a superhemispherical cubic-zirconia solid immersion lens. See section IA1 of the supplementary materials for a full breakdown of the heterostructure.

\subsection{Optics}
We use a confocal microscope with crossed polarisers on excitation and collection arms to excite the QD resonantly, and collect its emission. We excite the QD with circularly polarized light by using a quarter wave plate between the polarisers. The collected emission is spectrally filtered with an optical grating with a $20$GHz passband before being sent to a superconducting nanowire single photon detector (SNSPD, Quantum Opus One).

Two lasers are required for our experiments. The first--- a New Focus Velocity laser diode--- is resonant with the $\ket{\downarrow}\leftrightarrow\ket{\Downarrow\uparrow\downarrow}$ transition and is used for spin pumping for electron spin readout/reset. The second is used for electron spin control via a two-photon stimulated Raman process \cite{Bodey2019} and is based on a Toptica DL Pro laser diode fed through a Toptica BoosTA tapered amplifier. This Raman laser is $800$GHz-detuned from the trion excited state manifold. As required for Raman control, we derive two coherent laser fields from this single mode by feeding it through a fiber-based EOSPACE electro-optic amplitude modulator (EOM), which is driven with a microwave waveform. The resulting first-order sidebands after amplitude modulation are two coherent laser fields, separated by twice the microwave drive frequency, and whose relative phase is twice the phase of the microwave.

\subsection{Microwave signal}
Controlling the electron spin with a two-photon Raman process gives us effective microwave control over its Bloch vector. We can control the Rabi frequency, phase and detuning of the qubit drive by modifying respectively the power, phase and frequency of the EOM's microwave drive, all of which are imprinted onto the Raman beams by the EOM. An experimental sequence is thus defined by a microwave signal where all of these parameters, along with pulse timings, are set with a programmable microwave source. See section IA3 of the supplementary materials for the construction of the microwave sequences.\\

\textbf{Funding:} We acknowledge support from the US Office of Naval Research Global (N62909-19-1-2115), ERC PHOENICS (617985), EPSRC NQIT (EP/M013243/1), EU H2020 FET-Open project QLUSTER (862035) and EU H2020 Research and Innovation Programme under the Marie Sklodowska-Curie grant QUDOT-TECH (861097). Samples were grown in the EPSRC National Epitaxy Facility. D.A.G. acknowledges a St John's College Fellowship and C.LG. a Dorothy Hodgkin Royal Society Fellowship. \textbf{Author contributions:} D.A.G., C.L.G. and M.A. conceived and supervised the experiments. D.M.J. and U.H. implemented and carried out the experiments. D.M.J. and U.H. performed the data analysis. L.Z. and D.M.J. developed the theory and performed the simulations, with guidance from C.L.G. and D.A.G. E.C. and M.H. grew the material. All authors contributed to the discussion of the analysis and the results. All authors participated in preparing the manuscript. \textbf{Competing interests:} The authors declare that they have no competing interests. \textbf{Data and materials availability:} All data needed to evaluate the conclusions in the paper are present in the paper and the Supplementary Materials.

\bibliographystyle{ScienceAdvances.bst}
\bibliography{ramsey_cooling}

\end{document}


\title{Optimal purification of a spin ensemble by quantum-algorithmic feedback: Supplementary materials
}

\author{Daniel M.\,Jackson\textsuperscript{1,*}}
\author{Urs Haeusler\textsuperscript{1,*}}
\author{Leon Zaporski\textsuperscript{1}}
\author{Jonathan H.\,Bodey\textsuperscript{1}}
\author{Noah Shofer\textsuperscript{1}}
\author{Edmund Clarke\textsuperscript{2}}
\author{Maxime Hugues\textsuperscript{3}}
\author{Mete Atat\"ure\textsuperscript{1,$\dagger$}}
\author{Claire Le Gall\textsuperscript{1,$\dagger$}}
\author{Dorian A.\,Gangloff\textsuperscript{1,$\dagger$}}

\noaffiliation

\affiliation{Cavendish Laboratory, University of Cambridge, JJ Thomson Avenue, Cambridge, CB3 0HE, UK}
\affiliation{EPSRC National Epitaxy Facility, University of Sheffield, Broad Lane, Sheffield, S3 7HQ, UK}
\affiliation{Universit\'e C\^ote d'Azur, CNRS, CRHEA, rue Bernard Gregory, 06560 Valbonne, France
\\ \ \\
\textsuperscript{*}\,These authors contributed equally to this work.
\\
\textsuperscript{$\dagger$}\,Correspondence should be addressed to: ma424@cam.ac.uk; cl538@cam.ac.uk; dag50@cam.ac.uk.
\\ \ \\
}
\maketitle
\tableofcontents

\section{Experimental methods}
\label{sec_methods}
\subsection{Setup}
\label{sec_setup}
\subsubsection{Sample}
\label{sec_sample}

\begin{figure}
    \centering
    \includegraphics[width = \columnwidth]{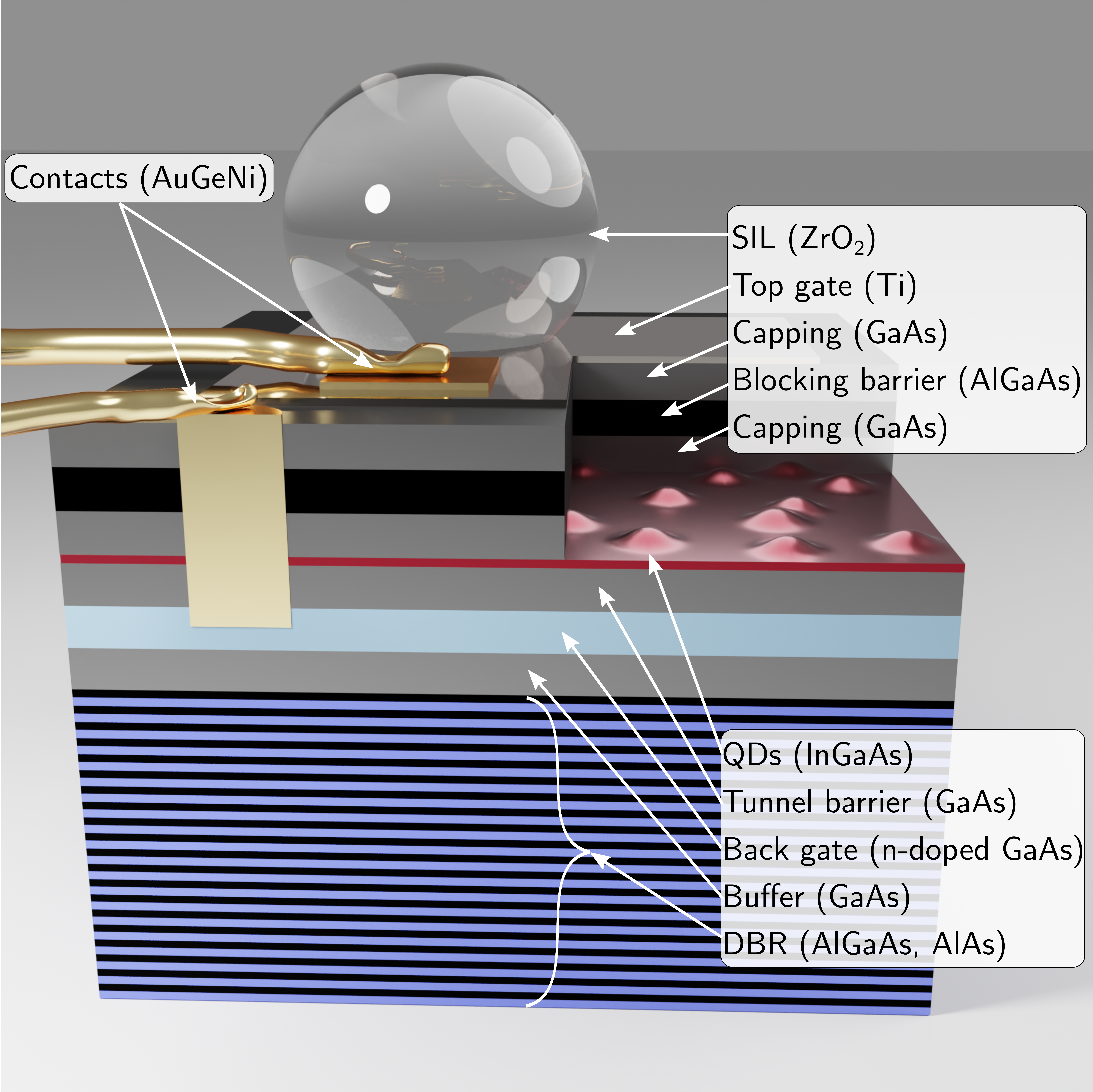}
    \caption{\textbf{Sample} Heterostructure of our QD sample (not to scale) with a cut-out above the QD layer. See section \ref{sec_sample} for a breakdown of the structure.}
    \label{fig_sample}
\end{figure}

The heterostructure of the wafer used in this work, which has been used in previous studies \cite{Stockill2016, Ethier-Majcher2017, Huthmacher2018, Bodey2019, Gangloff2020}, is depicted schematically in Fig.\,\ref{fig_sample}. The InGaAs QD layer (shown in red), is capped above and below with GaAs (gray). The layer below is $35$\,nm deep and forms a tunnel barrier between the QD, and the Fermi sea of the n-doped GaAs back contact (light blue). The back contact combined with the semi-transparent titanium top gate ($6$\,nm) forms a Schottky diode structure which allows to control the charge state of the QD. An electron in the ground state is a stable configuration for a time $T_1=50\,\mu$s thanks to a tunnel barrier between the QD layer and the Fermi sea of the back contact. The two diode gates are electrically contacted with ohmic AuGeNi contacts (shown in gold). Above the top capping layer is a blocking barrier of AlGaAs (black) to prevent charge leakage, and then a final capping of GaAs. At the bottom of the heterostructure is a distributed Bragg reflector to improve photon emission from the top surface. The photon collection is further enhanced to 10\% at the first lens with a superhemispherical cubic-zirconia solid immersion lens.

\subsubsection{Optics}
\begin{figure*}
    \centering
    \includegraphics[width = \textwidth]{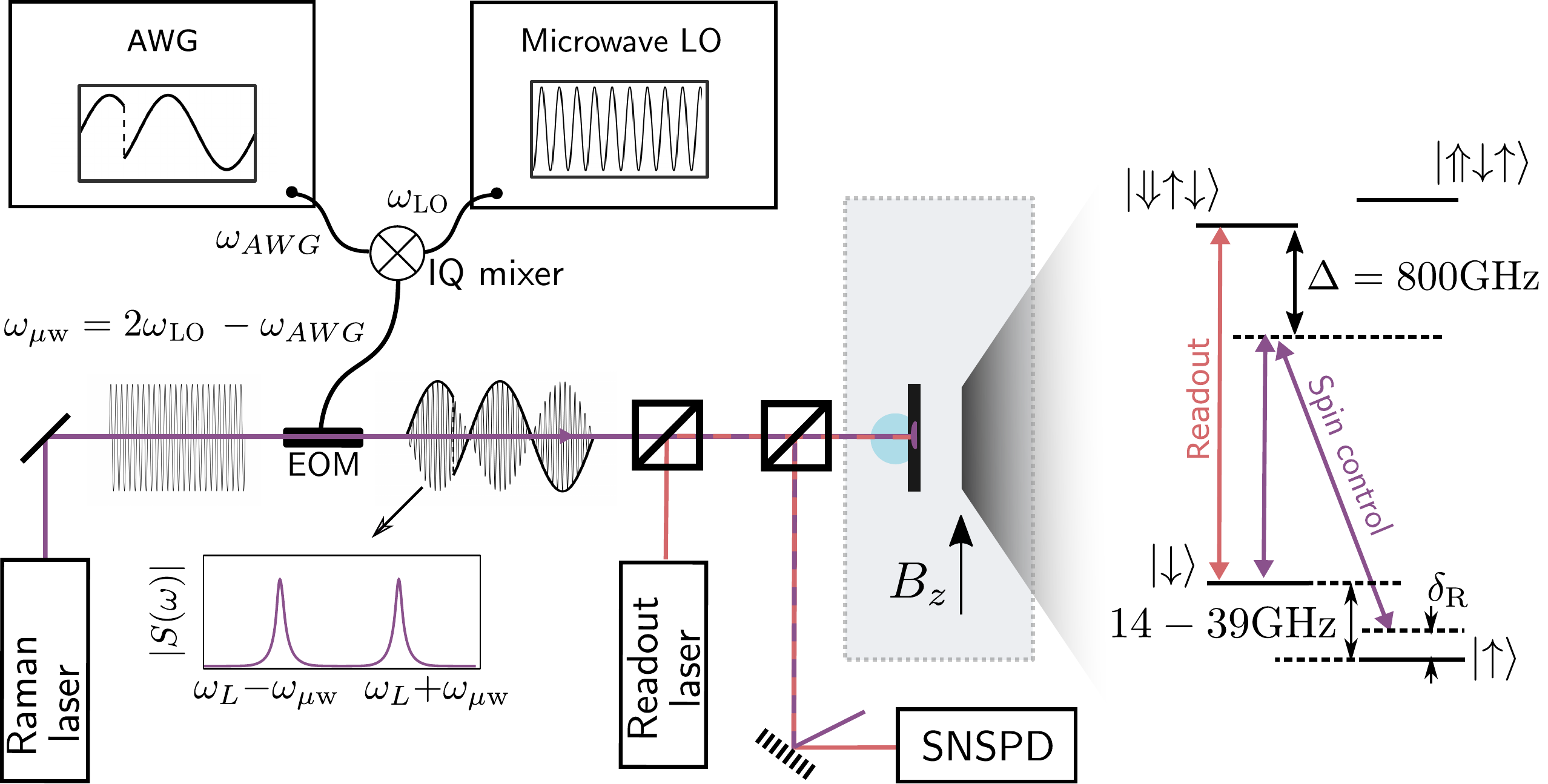}
    \caption{\textbf{Optical and microwave setup.} Adapted from ref\cite{Gangloff2020}. From left to right we have a Raman laser $800$\,GHz detuned from the trion manifold that is fed through an electro-optic modulator (EOM). The EOM is driven by a microwave signal derived from the IQ-mixing of an arbitrary waveform generator (AWG) with a local oscillator (LO). The EOM output is then 2 coherent laser fields. A beamsplitter combines the Raman laser with a resonant laser for readout, which is sent to the sample. The sample sits in a bath cryostat at $4$\,K, with a $3.5$\,T magnetic field applied in Voigt geometry. The QD emission is collected and excitation light is filtered by polarization and color before being counted on a superconducting-nanowire single-photon detector (SNSPD). The rightmost section depicts the QD energy level diagram and excitation laser frequencies.}
    \label{fig_optics}
\end{figure*}

We use a confocal microscope with crossed polarisers on excitation and collection arms to excite the QD resonantly, and collect its emission. We excite the QD with circularly polarized light by using a quarter wave plate between the polarisers. The collected emission is spectrally filtered with an optical grating with a $20$\,GHz passband before being sent to a superconducting nanowire single photon detector (SNSPD, Quantum Opus One).

Two lasers are required for our experiments. The first--- a New Focus Velocity laser diode--- is resonant with the $\ket{\downarrow}\leftrightarrow\ket{\Downarrow\uparrow\downarrow}$ transition and is used for spin pumping and electron spin readout/reset. The second is used for electron spin control via a two-photon stimulated Raman process \cite{Bodey2019} and is generated by a Toptica DL Pro laser diode fed through a Toptica BoosTA tapered amplifier. This Raman laser is $800$\,GHz detuned from the trion excited state manifold. As required for Raman control, we derive two coherent laser fields from this single mode by feeding it through a fiber-based EOSPACE electro-optic amplitude modulator (EOM), which is driven with a microwave waveform (section \ref{sec_microwave}). The resulting first-order sidebands after amplitude modulation are two coherent laser fields, separated by twice the microwave drive frequency $\omega_{\mu \text{w}}$, and whose relative phase is twice the phase of the microwave \cite{Bodey2019}. With these we can drive the electron-spin resonance (ESR) at a frequency $\omega=2\omega_{\mu \text{w}}$.

\subsubsection{Microwave}
\label{sec_microwave}

Controlling the electron spin with a two-photon Raman process gives us effective microwave control over its Bloch vector. We can control the Rabi frequency, phase and detuning of the qubit drive by modifying respectively the power, phase and frequency of the EOM's microwave drive, all of which are imprinted onto the Raman beams by the EOM. An experimental sequence is thus defined by a microwave signal where all of these parameters, along with pulse timing and duration, are set programmatically with an arbitrary waveform generator (Tektronix AWG70001A), at a sampling rate of $6$ GSamples/s. We use 2 channels of the AWG to produce the I and Q components of this signal, which has a carrier frequency of $\omega_\text{AWG}=600$\,MHz. With these we can perform single-sideband mixing with a frequency-doubled local oscillator (LO) of frequency $\omega_\text{LO}\in[3.665, 10.065]$\,GHz to up-convert to the final microwave frequency $\omega_{\mu\text{w}}=2\omega_\text{LO}-\omega_\text{AWG}$. The IQ-mixer, which is an Analog Devices ADRF6780 board, handles internally the frequency-doubling of the LO, which is derived from a Rohde \& Schwarz SMF100A source.

\subsection{Techniques}
\label{sec_techniques}
\subsubsection{Full experimental pulse sequence}
\label{sec_pulse_seq}

\begin{figure*}
    \centering
    \includegraphics[width = \textwidth]{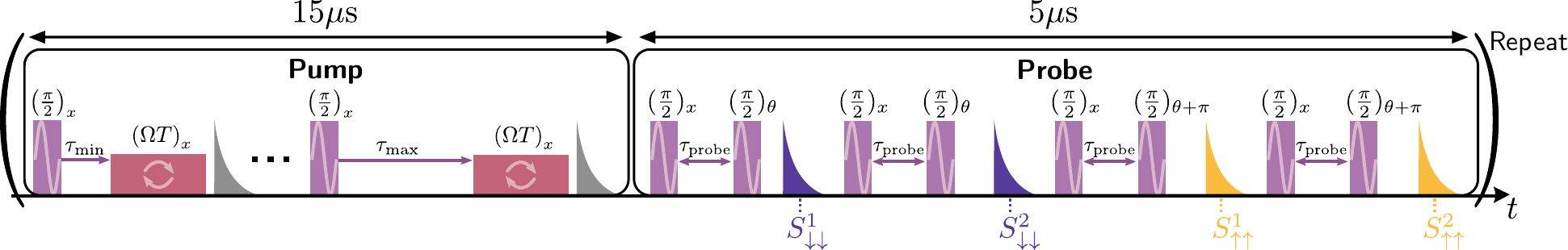}
    \caption{\textbf{Experimental pulse sequence.} A schematic of the ESR control pulses (pink and red square pulses) and readouts (gray, purple and yellow spin-pumping transients) that make up a single repeating block of the entire experiment (timings not to scale). The purple and yellow spin pumping transients correspond to reading out opposite electronic spin populations, achieved with a relative $\pi$ phase shift on the second Ramsey gate $(\frac{\pi}{2})_{\theta+\pi}$. The phase of the second Ramsey gate is serrodyned at a frequency $\omega_\text{serr}$ via $\theta=2\pi\tau_\text{probe}\omega_\text{serr}$.}
    \label{fig_sequence}
\end{figure*}

Figure \ref{fig_sequence} depicts the pulse sequence that makes up a single repeating block of the experiment. This block repeats at a rate of $50$\,kHz and is composed of a $15\,\mu$s pump section, where we run the feedback algorithm to prepare a target state of the nuclear ensemble, and a $5\,\mu$s probe section, where we measure the resulting electronic free-induction decay (FID) to extract the probability distribution $p(A_\text{c}\Delta I_z)$. The pump section is described in detail in the main manuscript. The probe section consists of $4$ separate Ramsey interferometry measurements, two of which probe the spin-$\uparrow$ population, $\rho^\text{e}_{\uparrow\uparrow}$, and two of which probe $\rho^\text{e}_{\downarrow\downarrow}$. Having two of each is not necessary, and is only done to improve the signal-to-noise ratio.

Selecting which spin population we readout is done as follows: the readout laser is always resonant with the $\ket{\downarrow}\leftrightarrow\ket{\Downarrow\uparrow\downarrow}$ transition making $\ket{\downarrow}$ the bright state. Thus we select which population to readout by choosing to swap the electron spin populations with a $\pi$-pulse before optical pumping, yielding a spin fluorescence signal, $S_{\downarrow\downarrow}$ or $S_{\uparrow\uparrow}$, proportional to the population, $\rho^\text{e}_{\downarrow\downarrow}$ or $\rho^\text{e}_{\uparrow\uparrow}$ respectively. In practice we replace this additional $\pi$-pulse with a $\pi$-phase on the second Ramsey gate, which achieves the same result. This avoids an erroneous disparity in the two populations that would result from the addition of a pulse with finite fidelity. Note that $S$ is the integrated fluorescence over the entire spin-pumping transient after background subtraction. We may then calculate the average spin populations:

\begin{equation}
\rho^e_{\uparrow\uparrow}=\Big\langle\frac{S^1_{\uparrow\uparrow}+S^2_{\uparrow\uparrow}}{S^1_{\uparrow\uparrow}+S^2_{\uparrow\uparrow}+S^1_{\downarrow\downarrow}+S^2_{\downarrow\downarrow}}\Big\rangle,
\label{equ_populations}
\end{equation}
where we combine the two repeated spin readouts, $1$ and $2$, for improved signal-to-noise ratio, and we average over the many repetitions made during a given integration time. Finally we note that each of the $4$ Ramsey interferometry measurements has a phase $\theta=2\pi\tau_\text{probe}\omega_\text{serr}$ added to the second $(\frac{\pi}{2})$-gate, where $\omega_\text{serr}$ is a serrodyne frequency. This adds a Fourier component at frequency $\omega_\text{serr}$ to the FID making fitting the decay envelope robust against a few-MHz systematic detuning arising from the optical Stark shift during the Ramsey gates \cite{Jackson2021}.

\subsubsection{Polarization protocol}
\label{sec_pol}

In the main text we describe how we can polarize the QD nuclear ensemble by stepping in time the ESR drive detuning, $\delta(t) = \omega_\text{e}-\omega= -A_\text{c}I_z^\text{lock}$. We achieve this by varying the LO frequency in discrete steps $\Delta\omega_\text{LO}$ resulting in steps of detuning $\Delta\delta=-4\Delta\omega_\text{LO}$. In this way we step the detuning by $20$\,MHz every $\sim50$\,ms, amounting to $\sim2500$ repeats of the entire experimental sequence (sec. \ref{sec_pulse_seq}) per step ensuring that the nuclear spin system reaches steady-state at every step. As we polarize the nuclei, the resulting Overhuaser shift alters the electron spin splitting and thus the optical transition frequency to the trion manifold. In order to polarize beyond the trion linewidth we therefore need to compensate this effect with a DC Stark shift. We step the Schottky diode gate bias with the LO frequency to maintain single-photon resonance with the fixed-frequency readout laser.

\section{Additional notes on data analysis}
\label{sec_add_analysis}

\subsection{Entropy}
\label{sec_entropy}
In the main text, the concept of entropy was applied to quantify the purity of an arbitrary distribution. To this end, we employed the limiting density of discrete points (LDDP), which is an extension of Shannon entropy to continuous probability distributions. It is defined by
\begin{align*}
    H_N(X) &= \log(N) + H(X) \\
    \text{with} \quad H(X) &= - \int p(x) \log \frac{p(x)}{m(x)}\, dx,
\end{align*}
where $N$ is the number of points discretising the continuous distribution $p(x)$, and $m(x)$ is an invariant measure of the density of points as $N \rightarrow \infty$.

In our case, we evaluated the probability distribution $p(A_\mathrm{c} \Delta I_z)$ in the range from $-250\,\mathrm{MHz}$ to $250\,\mathrm{MHz}$, hence $m(x) = 2\,\mathrm{GHz^{-1}}$. We chose $N=400$, which leads to a constant offset of $\log 400 = 6$.

\subsection{Calculating number of nuclei}
\label{sec_N}

We can extract the number of nuclei in the QD from a measurement of the hyperfine constant per nucleus, $A_\text{c}$, and $T_2^*$ at thermal equilibrium at infinite temperature. The former is provided by the ESR-shift measurement \cite{Jackson2021} reported in Fig.\,3c of the main text, and the latter from the FID in Fig.\,2c of the main text. This FID was measured at thermal equilibrium at $4$\,K, which is effectively infinite temperature when comparing to the relevant nuclear Zeeman energy scale. This means we can safely assume that the nuclear state is fully mixed. In this case the nuclear probability distribution is a Gaussian, $p(\Delta I_z)=e^{-\frac{\Delta I_z}{2\sigma^2}}$, with standard deviation given simply by

\begin{align*}
    \sigma^2 = \langle I_z^2 \rangle =\biggl<\biggl(\sum_i^N I_{i,z}\biggr)^2\biggr> = \sum_i^N \langle I_{i,z}^2 \rangle =\frac{1}{3}NI_j(I_j+1),
\end{align*}
where $i$ indexes an individual spin and we have assumed $\langle I_z \rangle=0$. Taking a single species with $I_j=3/2$ gives $\sigma=\sqrt{5N/4}$. Since $p(\Delta I_z)$ and the FID are related by a Fourier transform, the FID is also Gaussian with $T_2^*=1/\sqrt{2}\pi A_\text{c}\sigma$. Using $T_2^*=1.52(5)$\,ns and $A_\text{c}=0.63(2)$\,MHz we can then calculate $N=49(4)\cdot 10^3$.

One can also extract an estimate of $N$ from $T_2^*$ by using the hyperfine constants of the material and assuming an indium concentration \cite{Stockill2016}. Since indium is the species with the highest spin and the largest hyperfine constant, its concentration has a significant effect on $N$, which can range from $48,000$ to $110,000$ for concentrations from $0.2$ to $0.7$. As such our estimate of $N=49,000$ is entirely reasonable.

\section{Additional data}
\label{sec_add_data}
\subsection{Minimum feedback capture range}
\label{sec_capture_range}

\begin{figure}
    \centering
    \includegraphics[width = \columnwidth]{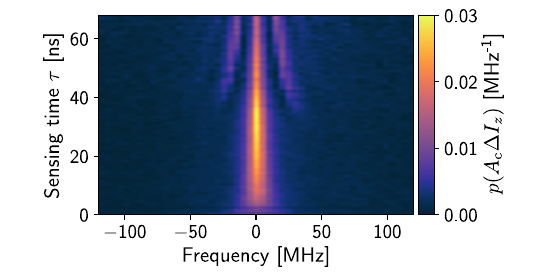}
    \caption{\textbf{Minimum capture range.} 2D plot of $p(A_\text{c}\Delta I_z)$ versus frequency, $A_\text{c}\Delta I_z$, and sensing time $\tau$. In this experiment we do not dynamically sweep the sensing time, but rather use a constant value for all $44$ repeats (per preparation cycle) of the algorithm. For sufficiently small $\tau\lesssim 35$\,ns we avoid multistability and prepare a single-mode distribution.}
    \label{fig_min_cap_range}
\end{figure}

As detailed in the main manuscript, if we seek to purify the nuclear ensemble to a single mode we must start feedback with a sufficiently broad capture range as not to populate the next-nearest stable points, Fig.\,1c. Given the feedback bandwidth of $\sim 20$\,MHz, determined by the width of the HH resonance, we expect that the broadest capture range we require is $1/\tau_\text{min} \sim 20$\,MHz or equivalently $\tau_\text{min}\sim 50$\,ns. In this way, at the beginning of the feedback preparation sequence the next nearest stable points are beyond the bandwidth of the feedback and thus are very weak attractors ensuring purification to a single mode. In Fig.\,\ref{fig_min_cap_range} we plot the results of an identical experiment to Fig.\,3c of the main manuscript, i.e feedback with a single sensing time, but for small $\tau$. From this we can see a region of $\tau\lesssim 35$\,ns where it is not possible to populate neighboring stable points which justifies using $\tau_\text{min}=30$\,ns throughout our experiments.

\section{Feedback function formalism}
\label{sec_feedback_func}

With the spin bath evolving under drive and dissipation we can construct semiclassically a rate equation that captures the evolution of the mean value $\langle I_z \rangle$ valid over a coarse-grained timescale much larger than the time of a single algorithm cycle \cite{Yang2013, Hogele2012}. We stress that the fully quantum approach used in our modeling (section \ref{sec_model}) is the most complete, but we can nevertheless gain intuition about the feedback with this method. The algorithm itself results in directional $\langle I_z \rangle$-changing rates $W_\pm(\langle \Delta I_z \rangle)$ that stabilize the value of $\langle I_z \rangle$. Further, there will be relaxation processes in any central spin system that lead to spin diffusion, either intrinsic or electron-mediated, which re-thermalizes the bath at a rate $\Gamma_\text{d}$. For example, in our QD system electron-mediated nuclear-nuclear spin interactions \cite{Latta2011, Wust2016} lead to diffusion at a rate $\mathcal{O}(10\,\text{Hz})$ \cite{Gangloff2020}. The rate equation then reads
\begin{equation}
    \frac{d\langle I_z \rangle }{dt} = W_+(\langle \Delta I_z \rangle)-W_-( \langle \Delta I_z \rangle)-\Gamma_\text{d} \langle I_z \rangle,
\end{equation}

The directional rates can be constructed intuitively, since we will drive one spin flip per algorithm cycle in response to a $\Delta I_z$ fluctuation provided $\tau=1/4A_0\Delta I_z$. Assuming unitarity in the flip-flop interaction and that one cycle takes $\tau+1/2A_\text{ff}$, neglecting the spin-pumping time, we have

\begin{equation}
    W_\pm(\langle \Delta I_z \rangle )=\frac{1\mp \sin(2\pi A_0\langle \Delta I_z \rangle \tau)}{2\tau+1/A_\text{ff}}.
\end{equation}
The feedback curve then becomes
\begin{equation}
\frac{d\langle I_z \rangle }{dt} =\frac{-\sin(2\pi A_0\langle \Delta I_z \rangle \tau)}{\tau+1/2A_\text{ff}} -\Gamma_\text{d} \langle I_z \rangle,
\label{equ_dizdt}
\end{equation}
which we report in the main manuscript with $\Gamma_\text{d}\ll 1/\tau +1/2A_\text{ff}$.

\section{Modeling}
\label{sec_model}

\subsection{Outline}
\label{sec_outline}

We seek to model the coupled electron-nuclear dynamics in our QD system during the feedback algorithm. From the resulting complete density matrix we may take a partial trace over the electron to leave only the density matrix of the nuclear spin system. From this we calculate $p(I_z)$ and compare to our experimental data via either of our feedback metrics: entropy or $T_2^*$.

The nuclear ensemble comprises three species: arsenic ($I=3/2$), indium ($I=9/2$) and gallium ($I=3/2$), but we may safely neglect gallium's contribution to the feedback dynamics owing to a much weaker quadrupolar contribution \cite{Stockill2016}. To reduce the computational complexity of our model we make our first approximation: we run two independent simulations for arsenic and indium separately and average the results. In this way we avoid squaring the dimensionality of the nuclear Hilbert space and its associated complexity.

Our second simplifying approximation is to model the nuclei as an ensemble of $N$ spin-$1/2$ particles, with a uniform hyperfine coupling to the electron. This allows to parameterize the nuclear state by the set of quantum numbers ${I,I_z}$, where the integers $I$ and $I_z$ range from $0$ to $N/2$ and $-I$ to $I$, respectively. In this way we reduce the calculation on the full $2\times2^N$-dimensional Hilbert space to $N/2$ calculations on $2\times(2I+1)$-dimensional independently evolving, uncorrelated Hilbert spaces. In this basis the full electron-nuclear density operator can be written as
\begin{equation}
\rho = \sum_{S_z}\sum_{S_z^\prime}\sum_{I,I_z}\sum_{I^\prime,I_z^\prime}\rho_{S_z,S_z^\prime,I,I_z,I^\prime,I_z^\prime} \ketbra{S_z}{S_z^\prime}\otimes \ketbra{I,I_z}{I^\prime,I_z^\prime}.
\label{equ_rho}
\end{equation}

Solving for the driven-dissipative dynamics of the coupled electron-nuclear system via a Linblad master equation requires the use of superoperators of size $D^2\times D^2$, where $D$ is the Hilbert space dimensionality. In each $I$-manifold we have $D=2(2I+1)$, meaning that for realistic $N\sim 10^5$, the large-$I$ manifolds would require prohibitively large matrices of size $N^2\times N^2$ to represent the superoperators. To overcome this we make a second approximation to split the evolution into distinct unitary quantum operations and non-unitary dissipation and dephasing operations. This reduces the computational complexity to simple $D\times D$-matrix multiplication for both coherent processes, via exact diagonalisation, and incoherent processes, via Kraus operators.

Even after reducing the matrix size to $D\times D$, this is still prohibitively large for manifolds of $I\sim N\approx 10^5$, which is where the third approximation comes in: manifolds of very large $I$ can be neglected. This is because we weight the outcome of a simulation in a given $I$-manifold, $\rho_I$, by its degeneracy when calculating the final expectation value of any observable $A$:

\begin{equation}
\text{Tr} \rho A=\sum_{I} w_{I,N} \text{Tr} \rho_I A
\end{equation}.

We calculate these weights exactly in section \ref{sec_weights} but suffice to say that they are peaked strongly around $I=\sqrt{N/2}\approx160$, and decay exponentially with $I$ from there on. As such, assuming  $I$ is indeed distributed thermally according to these weights, we capture the vast majority of the dynamics by a coarse-grained simulation in only $46$ manifolds with $I$ values of $0,14,28,...,630$. In each of the manifolds we further truncate the Hilbert space by only simulating with $I_z$ values ranging from $-I/14$ to $I/14$, i.e: only around states of low polarisation, which is a good approximation given that our feedback procedure purifies the nuclear state very close to the zero-polarization macrostate.

\subsection{Unitary evolution during the algorithm}

During the sensing and actuate parts of step $j$ of the feedback algorithm we consider the unitary part of the evolution to act on the density operator in the usual way:
\begin{equation}\label{cooling_recursion}
\rho_{j+1}=U_j\rho_j U_j^\dagger 
\end{equation}

The engineered Hamiltonians during sensing and actuation are $H_\text{sense}=\delta S_z+\omega_\text{n}I_z+A_\text{c}S_zI_z+A_\text{nc}S_zI_x$ and $H_\text{act}=\Omega S_x+\omega_\text{n}I_z-\frac{A_\text{nc}}{4}(\Tilde{S}_+I_-+\Tilde{S}_-I_+)$ respectively, where $\Tilde{S}_\pm=S_z\pm iS_y$. Given these, the unitary evolution during step $j$ of the algorithm is generated by: 
\begin{equation}
U_j=\underbrace{e^{-iTH_\text{act}}}_{U_{\text{act}}}\underbrace{e^{-i\tau_j H_\text{sense}}}_{U_{\text{sense, j}}}\underbrace{\tfrac{1}{\sqrt{2}}(\mathds{1}-i\sigma_x)}_{R_x(\frac{\pi}{2})},
\end{equation}
where $R_i(\theta)$ represent rotations of angle $\theta$ about axis $i=x,y,z$, and $\sigma_i$ are the Pauli matrices.

The action of $U_\text{sense, j}R_x(\frac{\pi}{2})$ on the density operator expressed as in equation \ref{equ_rho} is straightforward, provided we assume $A_\text{nc}\ll A_\text{c}$ and neglect the $A_\text{nc}S_zI_x$ term -- we re-introduce the effect of $A_\text{nc}S_zI_x$ with a semi-classical approximation in section \ref{sec_non_unit}. In this case $H_\text{sense}$ is diagonal in the $\ket{S_z}\otimes \ket{I,I_z}$ basis and the matrix exponential of $U_\text{sense}R_x(\frac{\pi}{2})$ leads to a simple phase acquisition. The action of $U_{\mathrm{act}}$ is less straightforward to compute since $H_\text{act}$ is not diagonal. We first re-write as $U_{\mathrm{act}}=e^{-iTH'_\text{act}}R_{-y}(\frac{\pi}{2})$, where $H'_\text{act}=\Omega S_z+\omega_\text{n}I_z-\frac{A_\text{nc}}{4}(S_+I_-+S_-I_+)$, which is $2\times2$ block diagonal and can be diagonalized efficiently \cite{Kozlov2007}. The price we pay for this computational speed up is the implicit assumption that the electron-nuclear actuate gate is performed exclusively on Hartmann-Hahn (HH) resonance \cite{Hartmann1962, Henstra2008}, otherwise block diagonality is broken. To this end we model two nuclear species by running two simulations with nuclear Zeeman frequencies $\omega_n^\text{As}=25.3$\,MHz and $\omega_n^\text{In}=32.7$\,MHz, in each case imposing HH resonance, and average the resulting $p(I_z)$ distributions. To summarize this section, equation \ref{cooling_recursion} generates the exact unitary time evolution by simple matrix multiplication.

\subsubsection{Quantum circuit}
\label{sec_unitary}

This unitary part of the evolution, $U_j=U_\text{act}R_{-y}(\frac{\pi}{2})U_{\text{sense}, j}X_{\frac{\pi}{2}}$, is depicted as a quantum circuit in Fig.\,1b of the main text. Specifically we depict the action of $U_j$, where $\tau_j=1/4A_\text{c}$, on a pure nuclear state in the manifold $\{\ket{I, I_z}, \ket{I, I_z\pm1}\}$. In this way $U_{\text{sense}, j}$ effectively becomes a $Z$ rotation of the electron conditional on the nuclear polarization fluctuating one unit away from the lockpoint. Furthermore, by choosing a drive time $T=2/A_\text{nc}\sqrt{I(I+1)}$, the action of the $U_\text{act}$ gate becomes an exact SWAP operation (around zero polarisation, $I_z=0$). These specific choices mean the quantum circuit depicts the limit cycle behavior at the ultimate limit of the cooling algorithm: fluctuations of a single unit are detected with a conditional electron rotation and corrected deterministically with a SWAP operation.

\subsection{Adding non-unitary evolution}
\label{sec_non_unit}
There are several non-unitary processes to include in our model, the first of which is necessary for the feedback algorithm, and the remainder are dephasing and relaxation processes that hamper feedback.

\subsubsection{Electron reset}
\label{sec_reset}

This non-unitary process crucial to the operation of feedback resets the state of the electron spin via incoherent spin pumping. We simulate this process with the following amplitude-damping channel\cite{nielsen_chuang_2019} on the electron spin:
\begin{equation}
\mathcal{K}^\text{reset}: \rho \to K^\text{r}_0 \rho K^{\text{r}\dagger}_0 + K^\text{r}_1 \rho K^{\text{r}\dagger}_1,
\end{equation}
where the Kraus operators are defined as:
\begin{equation}
K^\text{r}_0=\begin{pmatrix}
1 & 0 \\ 0 & 0
\end{pmatrix} \otimes \mathds{1}_{\mathrm{nuc}}, \quad 
K^\text{r}_1=\begin{pmatrix}
0 & 1 \\ 0 & 0
\end{pmatrix} \otimes \mathds{1}_{\mathrm{nuc}}
\end{equation}
and we have safely assumed unit fidelity spin pumping \cite{atature_quantum-dot_2006}

\subsubsection{Transverse nuclear noise}
\label{sec_trans_noise}

In section \ref{sec_unitary} we neglected the non-colinear term $A_\text{nc}$ during sensing since including its effect explicitly in the $\ket{S_z}\otimes \ket{I,I_z}$ is computationally inefficient. Instead, we capture its effect on the dynamics by assuming  $I_x$ to be a zero-mean, classical Gaussian random variable $\mathcal{I}_x(t)$. This time-fluctuating transverse nuclear polarization induces broadening of the electronic energy levels, which results in electronic decoherence, that limits the cooling efficiency. 

Within this model, the post-sensing density operator is given by:
\begin{equation}
\rho(\tau)= \langle U(\tau) \rho(0)  U(\tau)^\dagger \rangle ,
\end{equation}
where 
\begin{equation}
    U(\tau)=e^{-i\tau A_\text{c}I_zS_z -iA_\text{nc}\int_0^\tau \,dt^\prime \mathcal{I}_x(dt^\prime)S_z}
\end{equation}
and the averaging $\langle... \rangle$ is done over all noise realizations through a path integral that commutes with the other operators in the evolution equation. The net effect of the evolution can be viewed simply as a coherence build-up due to the longitudinal polarisation, $A_\text{c}I_z$, alongside decoherence due to the transverse noise, $A_\text{nc}\mathcal{I}_x$. Commuting with the collinear-term, the noise imposes a transfer function $W(\tau)$ on electronic coherences\cite{Cywinski2008}, which we will now calculate. 

We start from writing this transfer function explicitly:
\begin{equation}\label{definition}
W(\tau)= \langle e^{-i\int_0^\tau \, dt^\prime A_\text{nc}\mathcal{I}_x(t^\prime)} \rangle
\end{equation}
Since the transverse polarization is a Gaussian random variable, the sum over all its realizations in time - labeled as $X$ - is also a Gaussian random variable:
\begin{equation}
W(\tau)= \int_{-\infty}^\infty \, dX \frac{1}{\sqrt{2\pi}\sigma_\tau} e^{-\frac{X^2}{2\sigma_\tau^2}}e^{iX}=e^{-\sigma_\tau^2/2}
\end{equation}
We can find $\sigma_\tau$ by taking (without loss of generality) $\langle X \rangle = 0$ and evaluating the auto-correlation function:
\begin{equation}
\sigma_\tau^2 =\langle X^2(\tau) \rangle = A_{nc}^2 \int_0^\tau dt_1 \int_0^\tau dt_2 \langle \mathcal{I}_x(t_1)\mathcal{I}_x(t_2) \rangle 
\end{equation}
Assuming that the noise is stationary, this correlator is dependent only on $T=|t_1-t_2|$, such that
\begin{equation}
\langle \mathcal{I}_x(t_1)\mathcal{I}_x(t_2)\rangle = \langle \mathcal{I}_x(T)\mathcal{I}_x(0) \rangle \quad \forall t_1,t_2.
\end{equation} 



\noindent In line with previous works \cite{Bluhm2011, Botzem2016} we take the amplitude of this classical variable to be equal to the correlator of our simplified model:
\begin{equation}
\langle{\mathcal{I}_x(T)\mathcal{I}_x(0)} \rangle=e^{-\Gamma T/2}\cos(\omega_n T)\langle{I_x(0)^2}\rangle,
\label{equ_correlator}
\end{equation} 
where $\Gamma$ is the rate of pure nuclear dephasing. Assuming $I_x(0)^2=I_y(0)^2$, equation \ref{equ_correlator} is equal to
\begin{equation}
\langle{\mathcal{I}_x(T)\mathcal{I}_x(0)} \rangle=\frac{1}{2}e^{-\Gamma T/2}\cos(\omega_n T)(I^2-\langle{I_z(0)^2}\rangle).
\end{equation}

Putting this all together, we have:
\begin{widetext}

\begin{align}
\sigma_\tau^2&=\frac{1}{2}(I^2-\langle{I_z(0)^2}\rangle A^2_{nc}\int_0^\tau dt_1 \int_0^\tau dt_2e^{-\Gamma|t_1-t_2|/2}\cos \omega_n(t_1-t_2) \\
&=(I^2-\langle{I_z(0)^2}\rangle A^2_{nc}\Bigg[\frac{\frac{\Gamma}{2}\tau}{\frac{\Gamma^2}{4} +\omega_n^2}-\frac{\Gamma \omega_n e^{-\Gamma \tau /2}\sin \omega_n \tau+(\frac{\Gamma^2}{4}-\omega_n^2)(1-e^{-\Gamma \tau /2}\cos \omega_n \tau)}{(\frac{\Gamma^2}{4}+\omega_n^2)^2} \Bigg],
\label{equ_w}
\end{align}
\end{widetext}
where the second line is the result of a change of variables $T=t_1-t_2$ such that $dt_2=-dT$, and a straightforward double integral. Since we have access to $I$ and $\langle{I_z(0)^2}\rangle$ throughout the simulations, we have everything required to calculate $W(\tau)=e^{-\sigma_\tau^2/2}$. Intuitively, we can see from equation \ref{equ_w} that at short times one observes some revivals of coherence related to Larmor precession, and at long times the decay of coherence becomes exponential.

Incorporating the above decay of electronic coherences into our calculation is achieved via the an electronic phase-damping channel:

\begin{equation}
\mathcal{K}^\text{TN}: \rho \to \sum_{i=0}^{2}{K^\text{TN}_i\rho K_i^{\text{TN}\dagger}},
\end{equation}
where
\begin{equation}
\begin{split}
K^\text{TN}_0&=\sqrt{W(\tau)}\mathds{1}_e\otimes \mathds{1}_n \\
K^\text{TN}_1 &= \sqrt{1-W(\tau)}\ketbra{\downarrow}{\uparrow}\otimes \mathds{1}_n \\
K^\text{TN}_2&=\sqrt{1-W(\tau)}\ketbra{\uparrow}{\downarrow}\otimes \mathds{1}_n
\end{split}
\end{equation}

\subsubsection{Optically induced electron relaxation}
\label{sec_opt_t1}

When shining non-resonant light on our QD system there exists an optically induced electron spin relaxation at a power-dependent rate $\Gamma_\text{opt}\propto \Omega$, likely the result of photo-activated charge noise in the device \cite{Bodey2019}. This relaxation has the largest effect during the actuate gate, where we illuminate the sample for a significant time. We incorporate this into the model with a generalized amplitude damping channel:

\begin{equation}
\mathcal{K}^\text{opt}: \rho \to \sum_{i=0}^{2}{K^\text{opt}_i\rho K_i^{\text{opt}\dagger}},
\end{equation}
where
\begin{equation}
\begin{split}
K^\text{opt}_0&=\begin{pmatrix}
1 & 0 \\ 0 & e^{\Gamma_\text{opt} T/2}
\end{pmatrix} \otimes \mathds{1}_{\mathrm{nuc}},\\
K^\text{opt}_1&=\begin{pmatrix}
e^{\Gamma_\text{opt} T/2} & 0 \\ 0 & 1
\end{pmatrix} \otimes \mathds{1}_{\mathrm{nuc}},\\
K^\text{opt}_2&=\begin{pmatrix}
0 & \sqrt{1-e^{\Gamma_\text{opt} T}} \\ \sqrt{1-e^{\Gamma_\text{opt} T}} & 0
\end{pmatrix} \otimes \mathds{1}_{\mathrm{nuc}}.
\end{split}
\end{equation}
We measure this relaxation rate via a separate Rabi driving measurement as per previous work \cite{Bodey2019}. For $\Omega=402$\,MHz we measure this relaxation rate to be $23.6$\,MHz, which we can re-scale linearly with Rabi frequency allowing us to fix $\Gamma_\text{opt}=1.7$\,MHz during actuation.

\subsubsection{Pure nuclear dephasing}
\label{sec_pd}
In section \ref{sec_trans_noise} we considered pure nuclear dephasing acting to damp transverse nuclear coherences that couple to the electron during sensing. The same nuclear dephasing processes, which arise from inhomogeneity in the quadrupolar coupling strength underpinning the non-colinear term and electron-mediated nuclear-nuclear spin coupling, are present during actuation. We include pure nuclear dephasing at a rate $\Gamma$, the same rate as in section \ref{sec_trans_noise}, via a nuclear phase damping channel in each $I$-manifold:

\begin{equation}
\mathcal{K}^\text{PD}: \rho \to \sum_{i=0}^{2I+1}{K^\text{PD}_i\rho K_i^{\text{PD}\dagger}},
\end{equation}
where
\begin{equation}
\begin{split}
K^\text{PD}_0&=e^{-\Gamma T/4}\mathds{1}_e\otimes \mathds{1}_n \\
K^\text{PD}_i &= \sqrt{1-e^{-\Gamma T/2}}\mathds{1}_e\otimes \ketbra{I, i-I-1},\ i\neq 0.
\end{split}
\end{equation}

\subsection{Combined evolution}
\label{sec_combined}

Having introduced all the individual ingredients that comprise the simulation, here we combine them into what constitutes a simulation of a single run of the feedback algorithm. To complete our quantum channel shorthand notation for the unitary gates, we define:
        
\begin{equation}
\begin{split}
\mathcal{U}^1_j: \rho &\to U_{\text{sense}, j}X_{\frac{\pi}{2}}\rho X_{\frac{\pi}{2}}^\dagger U_{\text{sense}, j}^\dagger\\
\mathcal{U}^2: \rho &\to R_{-y}(\frac{\pi}{2})\rho R_{-y}(\frac{\pi}{2})^\dagger \\
\mathcal{U}^3: \rho &\to U_{\text{act}}\rho U_{\text{act}}^\dagger.
\end{split}
\end{equation}
A simulation of a single run of the feedback algorithm then proceeds as:
\begin{equation}
\rho_{j+1} = \mathcal{K}^\text{reset}\mathcal{K}^\text{PD}\mathcal{U}^3\mathcal{K}^\text{opt}\mathcal{U}^2\mathcal{K}^\text{TN}\mathcal{U}^1_j  \rho_j
\label{equ_feedback_unit}
\end{equation}

As in the experiment, for our simulations we increase sensing time in each step, $\tau_j$, linearly from $\tau_0=30$\,ns to $\tau_{43}=\tau_\text{max}$ over $44$ runs of the fundamental feedback cycle (Eq. \ref{equ_feedback_unit}).

\subsection{Weighting I-manifolds}
\label{sec_weights}

We now have a complete description of how we simulate the feedback dynamics in a given $I$-manifold, the last consideration to make is how we weight the results from each manifold to obtain the quantity of interest, $p(I_z)$. The first step is to trace over the electron degrees of freedom as we are interested solely in the nuclear density matrix, as such $\rho$ refers to the nuclei only. We can re-express the $2^N\times 2^N$-dimensional total nuclear density operator in terms of the density matrices in individual, uncorrelated $I$-manifolds, $\rho_I$, as:
\begin{equation}
\rho = \bigoplus_{I=0}^{N/2} w_{I,N} \rho_I^{\oplus D_{I,N}}
\label{equ_rho_decomp}
\end{equation}
where $\oplus$ stands for matrix block concatenation and $D_{I,N}$ is the degeneracy of an $I$-manifold\cite{Dicke1954}:
\begin{equation}
D_{I,N}=\frac{N!(2I+1)}{(\frac{N}{2}-I)!(\frac{N}{2}+I+1)!}.
\end{equation} 
The weighting factors $w_{I,N}$ can be found by considering the $T=\infty$ initial state for the simulations:
\begin{equation}
\begin{split}
\rho_{T=\infty}&=\frac{1}{2^N}\mathds{1}_{2^N}\\
&=\frac{1}{2^N}\bigoplus_{I=0}^{N/2} \Bigg(\underbrace{\sum_{I_z=-I}^I\ketbra{I,I_z}}_{\mathds{1}_{2I+1}}\Bigg)^{\oplus D_{I,N}}\\
&=\frac{1}{2^N}\bigoplus_{I=0}^{N/2} (2I+1)\rho_{I,T=\infty}^{\oplus D_{I,N}}.
\end{split}
\label{equ_rho_inf}
\end{equation}
Comparing $\rho_{I, T=\infty}$ (Eq. \ref{equ_rho_inf}) to $\rho_{I}$ (Eq. \ref{equ_rho_decomp}), we obtain $w_{I,N}=(2I+1)/2^N$. With this we have everything required to calculate the expectation value of an arbitrary observable $A$ as:

\begin{equation}
\text{Tr} \rho A=\sum_{I=0}^{N/2} w_{I,N} D_{I,N} \text{Tr} \rho_I A,
\label{equ_expectation}
\end{equation}
which we use to extract the probability distribution of $I_z$:
\begin{equation}
p(I_z)= \sum_{I=0}^{N/2} w_{I,N} D_{I,N}\bra{I,I_z}\rho_I \ket{I,I_z}.
\end{equation}

The weights defined as  $w'_{I,N}=w_{I,N} D_{I,N}$ are difficult to compute on a desktop computer for $N> 30,000$, and so we use the approximate form\cite{Kozlov2007}:
\begin{equation}
    w'_{I,N}\approx \frac{2^\frac{5}{2}I(2I+1)}{\sqrt{\pi}N^\frac{3}{2}}e^\frac{-2I^2}{N}.
\end{equation}
In Fig.\,S2 we compare this approximate form (yellow curve) to the exact result (purple dotted curve) for $N=30,000$ where the quantitative agreement is clear. We further plot the approximate result for $N=49,000$, as well as the weights for the $46$ $I$-values used in the simulations (blue crosses) as discussed in section \ref{sec_outline}. Note than since we require the weights to sum to $1$ we re-normalize these $46$ values whilst preserving their relative weight.

\begin{figure}
    \centering
    \includegraphics[width = 0.8\columnwidth]{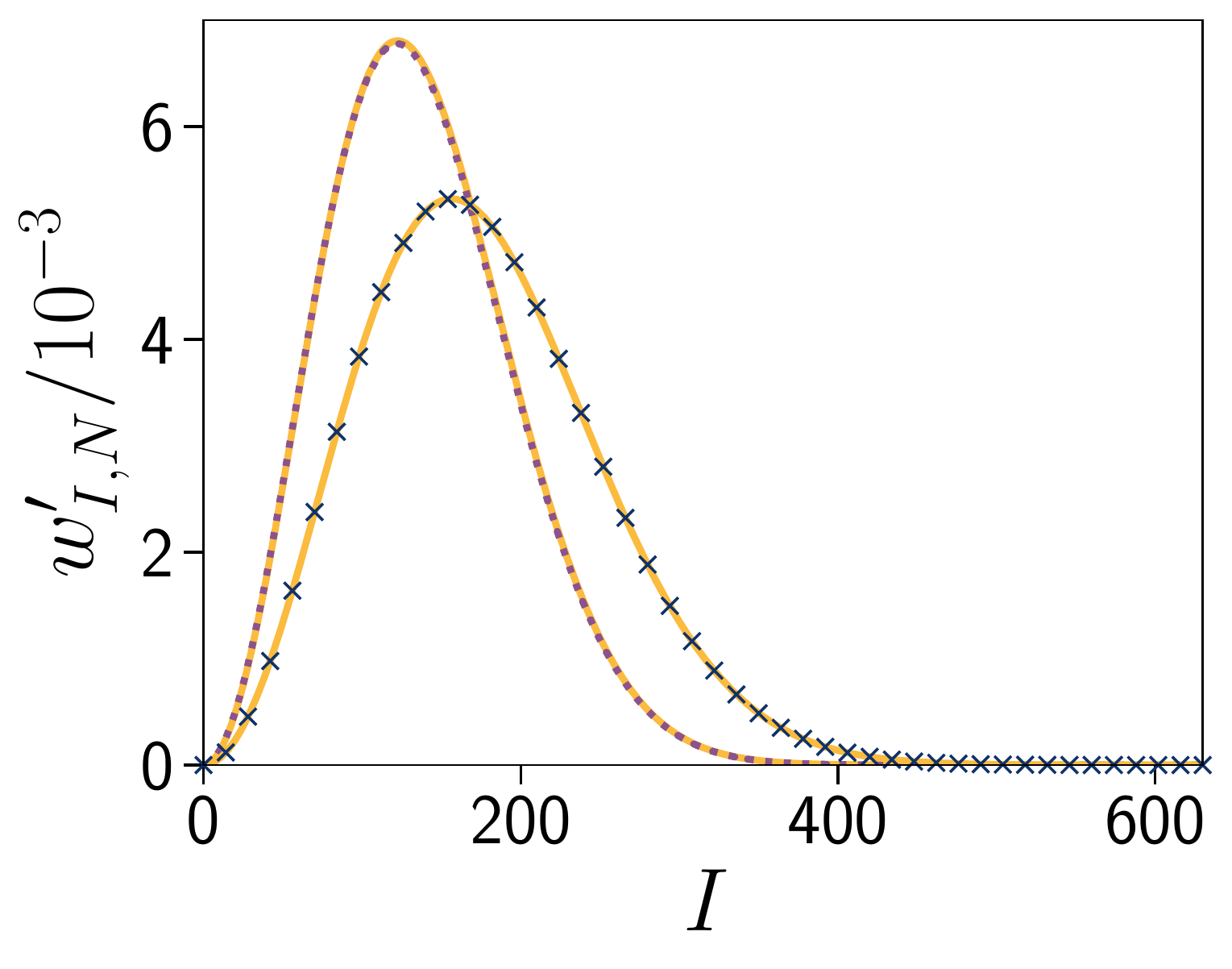}
    \caption{\textbf{Weighting factors for each $I$-manifold.} The purple dotted curve is the exact expression for the weights $w'_{I,N}$ for $N=30,000$, with the corresponding approximate version in yellow. We show only the latter for the case of $N=49,000$, overlaid with the $46$ sampled points used in our simulations (blue crosses).}
    \label{fig_weights}
\end{figure}

\subsection{Fitting model to data}
\label{sec_fitting}

There are four free parameters in our model: $A_\text{c}, A_\text{nc}$, $\Gamma$, and an additional parameter $\xi$ not yet discussed, which is the fraction of nuclei partaking in the magnon mode during actuation. This parameter is strongly motivated by previous work \cite{Gangloffeaaw2906, Jackson2021}, where it was inferred from modeling that only a small fraction of the total $N$ partake in coherent electron-nuclear exchange. This fraction $\xi$ is included ad-hoc as a scaling of $I\rightarrow \sqrt{\xi}I$ during the actuate gate, and thus scales the collective enhancement factor, $f(I,I_z)$, which combined with $A_\text{nc}$ sets the electron-nuclear coupling rate $\sim f(I,I_z)A_\text{nc}$. At the maximum degeneracy ($w_{I,N}$) point we have $f(I,I_z)\approx I = \sqrt{N/2}$, as such our scaling is conceptually equivalent to scaling $N\rightarrow\xi N$.

We get the first estimate of these four parameters from a least-squares fit of our model to the electron-nuclear exchange measurement of Fig.\,3c of the main text. Specifically we isolate the actuate step of our model, and extract the simulated ESR shift, $\Delta\omega = A_\text{c}\langle I_z \rangle$, versus $T$ as per equation \ref{equ_expectation}. Fitting this simulated curve to the data fixes precisely $A_\text{c}=0.63(2)$\,MHz and $\Gamma=6(2)$\,MHz, which we then use in our feedback model. However, since $\xi$ and $A_\text{nc}$ are heavily coupled, this measurement only constrains the product $\sqrt{\xi N}A_\text{nc}$. 

We then run our feedback simulation for each of the independent variables in the optimization curves of Figs.\,3a,c, extracting a $T_2^*$ value at each point. We then manually search the 2D space $\{A_\text{nc}, \xi\}$ to match the two simulated optimization curves to the two data sets simultaneously. There is an optimum in this space since $A_\text{nc}$ also strongly affects the transverse nuclear noise and in turn the attainable $T_2^*$ values. As such only one set of $\{A_\text{nc}, \xi\}$ reproduces simultaneously the coherent coupling rate and the scale of $T_2^*$. After this manual search we fix $\xi$ and perform a least squares fit to obtain our best estimate of $A_\text{nc}$ for this measurement. In conclusion, we obtain the fitted parameter set $A_\text{c}=0.63(2)$\,MHz, $\Gamma=6(2)$\,MHz, $\xi=0.42$ and very similar $A_\text{nc}$ values of $140(13)$\,kHz and $156(2)$\,kHz for the exchange measurement and $T_2^*$ optimization curves respectively.

\subsection{Modeling the ideal system}
\label{sec_ideal}
\begin{figure}
    \centering
    \includegraphics[width = \columnwidth]{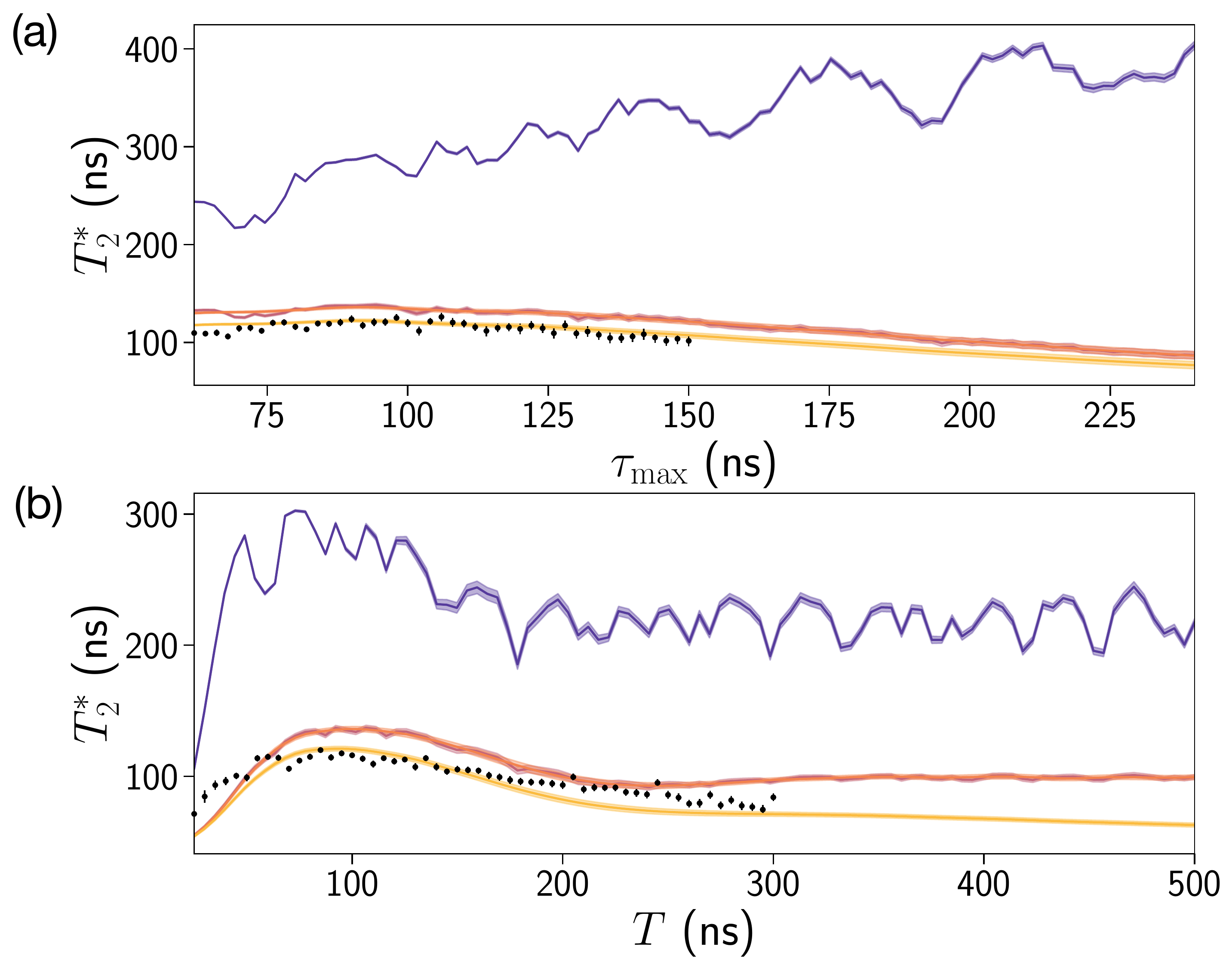}
    \caption{\textbf{Effects of relaxation and dephasing on feedback performance.} \textbf{(a, b)} $T_2^*$ versus maximum sensing time, $\tau_\text{max}$, and actuation time, $T$, respectively. Black circles are the measured values (also shown in the main text, Fig.\,3) with error bars representing the $68$\% confidence intervals. The solid curves are simulations of the data using the model parameters found in section \ref{sec_fitting}, but where we have removed different combinations of dephasing and relaxation mechanisms. The purple curve is an entirely unitary quantum evolution, except for the explicit dissipative reset step. For the pink curve we add in the electron dephasing due to transverse noise during sensing (section \ref{sec_trans_noise}). For the orange curve we further add pure nuclear dephasing during actuation (section \ref{sec_pd}). Lastly for the yellow curve we add the final mechanism, optically induced electron relaxation (section \ref{sec_opt_t1}).}
    \label{fig_ideal_model}
\end{figure}

The yellow curves of Fig.\,\ref{fig_ideal_model} are the result of our fitted simulation, extrapolated beyond the domain of the experimental data (black points). The other three colors represent simulations where we remove some or all of the sources of relaxation and dephasing in the model to see how the algorithm would perform in an idealized system.

Consider first the purple curves where we we simulate purely unitary evolution, i.e $\Gamma_\text{opt}=\Gamma=0$, and we neglect the transverse noise term $A_\text{nc}S_zI_x$. This latter consideration would be true of physical systems where a flip-flop actuator Hamiltonian could be engineered from a purely collinear hyperfine interaction. Strikingly we see oscillations in $T_2^*$ at the nuclear Zeeman frequency in both the $\tau_\text{max}$- and $T$-dependence, which is due to the buildup of transverse nuclear coherences during feedback which then oscillate in the magnetic field. Since we have two species with different gyromagnetic ratios, these oscillations beat. In the $T$-dependence it is clear that these high-frequency oscillations exist on top of a much lower frequency, low visibility oscillation corresponding to coherent electron-nuclear exchange. In the absence of any purification of $I$, as is the case in these simulations, the different exchange frequencies in different $I$ manifolds alone combine to heavily damp the exchange. As such, even in the absence of dephasing, we only observe the first maximum in $T_2^*$ versus $T$, which occurs at approximately $100$\,ns, in close agreement with the $\pi$-time measured in Fig.\,2d of the main text.
\begin{figure}
    \centering
    \includegraphics[width = \columnwidth]{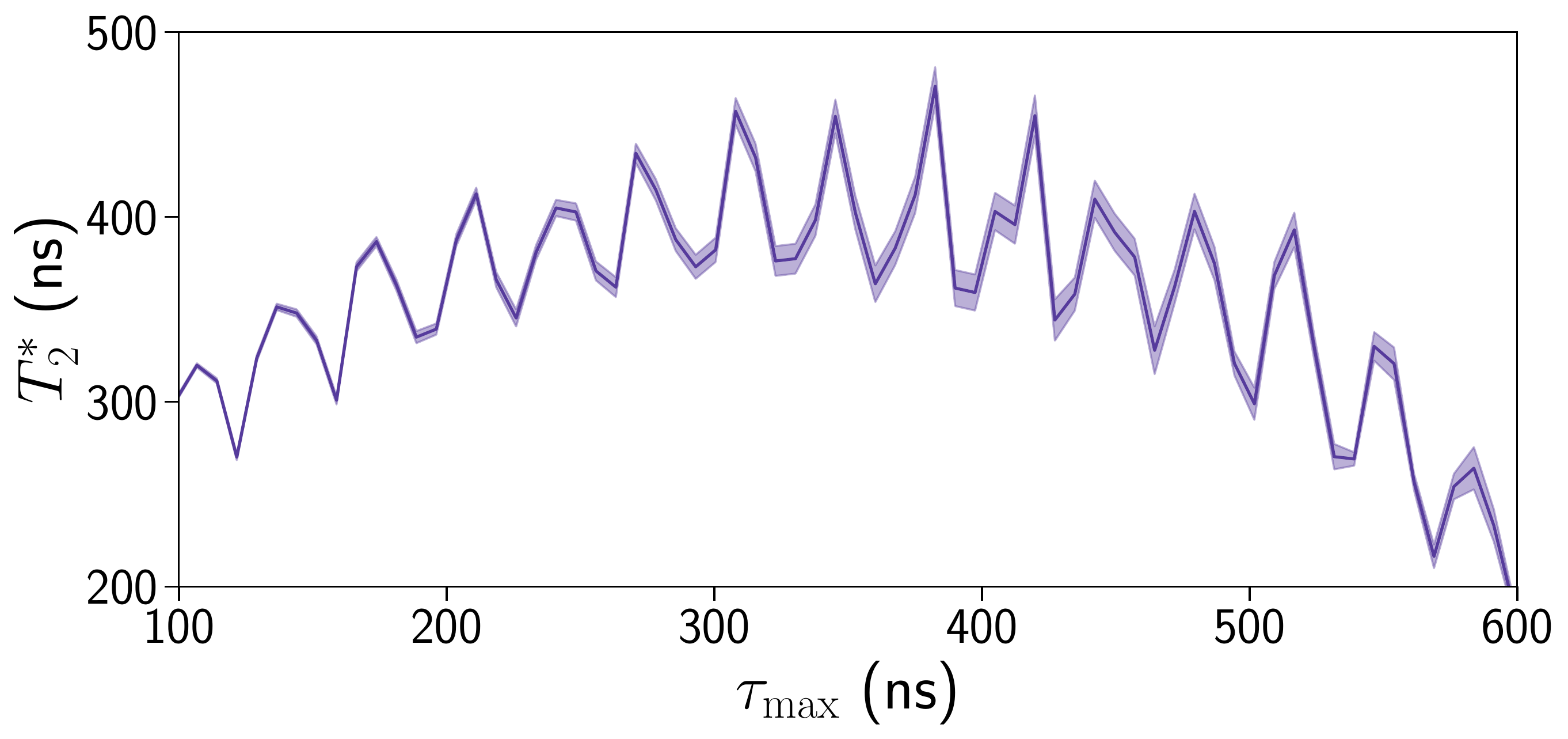}
    \caption{\textbf{$\tau_\text{max}$-dependence of feedback performance for a single-species unitary algorithm.} Simulated $T_2^*$ versus maximum sensing time,$\tau_\text{max}$, for entirely unitary quantum evolution during the feedback algorithm, except for the explicit dissipative reset step. Here we use the model parameters found in section \ref{sec_fitting}, but we only simulate a single nuclear species with Zeeman frequency $\omega_\text{n}=29$\,MHz.}
    \label{fig_unitary_single_spec}
\end{figure}
The behavior of the $\tau_\text{max}$-dependence in the unitary case is less clear in the presence of nuclear oscillations of two species so it is instructive to consider a simple single-species simulation, as in Fig.\,\ref{fig_unitary_single_spec}. Here it is clear that behind the nuclear oscillations there is a broad optimum in sensing time around $\tau_\text{max}\approx 350$\,ns. This is entirely consistent with our intuition that feedback is best when we are sensitive to single-spin fluctuations, namely $\tau_\text{max}=1/4A_\text{c}\approx 400$\,ns.

The process which has the most marked effect on the $\tau_\text{max}$-dependence is the dephasing introduced by transverse noise during sensing, which we add to the simulation in the pink curves of Fig.\,\ref{fig_ideal_model}. Optimum feedback is now a competition between minimising the effect of transverse noise and maximising sensitivity to spin fluctuations, reducing the optimal $\tau_\text{max}$ and the maximum $T_2^*$ attainable.

When we re-introduce dephasing, $\Gamma$, during the actuate step, transverse coherences get damped sufficiently to remove the remaining low-visibility oscillations at the nuclear Zeeman frequency, as seen in the orange curves. Finally re-introducing optically induced electron relaxation, $\Gamma_\text{opt}$, results in the best-fit curves in yellow. Its predominant effect is on the $T$-dependence, whereby it reduces the maximum $T_2^*$ achievable and slightly reduces the optimum exchange time thanks to driven diffusion as discussed in the main text.

\bibliographystyle{ScienceAdvances.bst}
\bibliography{ramsey_cooling_SI}